\documentclass[usenatbib,usegraphicx]{mn2e}

\def\Msun{\hbox{$\rm\thinspace M_{\odot}$}}
\def\Lsun{\hbox{$\rm\thinspace L_{\odot}$}}

\usepackage{subfigure}

\title{The Modelling of Feedback Processes in Cosmological Simulations of Disk
Galaxy Formation}

\author[F.~Piontek \& M.~Steinmetz]{
\parbox{14cm}{Franziska~Piontek and Matthias~Steinmetz}\\
Astrophysikalisches Institut Potsdam, An der Sternwarte 16, 14482 Potsdam, Germany}

\date{}
\pagerange{\pageref{firstpage}--\pageref{lastpage}}
\pubyear{2009}

\begin{document}
\label{firstpage}

\maketitle

\begin{abstract}
We present a systematic study of stellar feedback processes in simulations of
disk galaxy formation. Using a dark matter halo with properties similar to the
ones for the Milky Way's stellar halo, we perform a comparison of
different methods of distributing energy related to feedback processes to the
surrounding gas. A most promising standard model is applied to
halos spanning a range of masses in order to compare the results to disk
galaxy scaling relations. With few exceptions we find little or no angular
momentum deficiency for our galaxies and a good agreement with the angular
momentum-size relation. Our galaxies are in good agreement with the baryonic
Tully-Fisher relation and the slope of the photometric Tully-Fisher relation
is reproduced. We find a zero-point offset of 0.7 to 1 magnitudes, depending
on the employed IMF. We also study our standard feedback model in combination
with additional physical processes like a UV background, kinetic feedback, a
delayed energy deposition as expected for type Ia supernovae, mass return and
metal-dependent cooling. Only a combination of effects yields a real
improvement of the resulting galaxy by reducing the bulge, while including
metal-dependent cooling increases the bulge again. We find that in general the
stellar mass fraction of our galaxies is too high. In an ad-hoc experiment we
show that an removal of the bulge could reconcile this. However, the fit of the
Tully-Fisher relation can only be improved by delaying the star formation, but
not suppressing it completely. Our models do not seem to be efficient enough
to achieve either effect. We conclude that disk formation is a
complex, highly interconnected problem and we expect a solution to come from a combination of small effects.
\end{abstract}

\begin{keywords}
{galaxies: spiral - formation - evolution - structure - methods: \emph{N}-body
  simulations - hydrodynamics}
\end{keywords}

\section{Introduction}
The formation of galaxies is a complex process governed by gravitational
collapse and an energetic coupling between the
interstellar (and also intergalactic) gas and stellar evolution
processes. This includes the cooling of gas and subsequent formation of stars
as well as loss of mass and energy through stellar winds and supernova
explosions which is known as stellar feedback. Such feedback is essential for
the success of the widely accepted hierarchical model of galaxy formation
\citep{white78}. It regulates star formation by particularly preventing stars
from forming too early which would lead mostly to galaxies dominated by large,
old spheroids instead of disks \citep{white91}. This is inconsistent with
observations showing that up to 70\% of $\sim10^{12}~\rm{h}^{-1}~\Msun$ halos
host disk dominated late-type galaxies in the present day
universe \citep[e.g.][]{Weinmann2006,Park2007}. Feedback affects the thermodynamics of the interstellar medium (ISM)
in two ways. In supernova explosions,
large amounts of energy heat the surrounding gas and 
disrupt cold clouds, therefore efficiently quenching star formation. At the
same time, the surrounding gas is enriched with metals produced in stars
enabling more efficient cooling \citep{SD93} and therefore increasing star
formation. All of these processes
happen on scales of a few to tens of parsecs which is well below current
achievable resolution of cosmological simulations \citep[see however][]{ceverino09}. Therefore a range of so called
``subgrid models'' have been developed to model at least the effects on
larger, galactic scales (for example \citet{okamoto05},
\citet{Scannapieco2006}, \citet{SH2003}, \citet{stinson06}). Compared to early
work \citep[for example][]{Steinmetz1995,Navarro1997} and in combination with improved resolution, this has recently led to
considerable improvement in reproducing
individual realistic disk galaxies (see \citet{Abadi2003}, \citet{governato04},
\citet{governato07}, \citet{okamoto05}, \citet{Robertson2004},
\citet{scannapieco08}), though still none of them fulfilled all
characteristics of typical observed late-type spirals. Furthermore, none of
them resembled a bulge-less disk galaxy, despite of their non-negligible
numbers in the local universe \citep{kautsch}. The fundamental question of
disk galaxy formation therefore is still open.\par
All successful attempts to model disk galaxies which
have been achieved so far were simulated using handpicked dark
matter halos and carefully calibrated codes. Recent work by
\citet{scannapieco09} demonstrated that, when applying their model to a random
sample of halos, only a small fraction had a disk component and none was disk
dominated. Furthermore, \citet{scannapieco09} questions the usual assumption
that Hubble type and formation history are directly correlated. Considering the large
abundances of disk galaxies, this indicates that we may still lack an understanding
of the big picture. Furthermore, disks are susceptible to mergers which are
more common at high redshift and typically result in the formation of a
spheroid \citep{Peebles1982,Blumenthal1984,Davis1985,Gottlober2001,wechsler02,Cole2008,Wetzel2009}. One possible way out is to re-grow a
disk after a major merger, provided the merging progenitor disks are gas-rich
\citep{Robertson2006,Bullock2009}. A first example of this in a cosmological
simulation has been presented recently by \citet{governato08}, but again it is
an isolated case. \par
The work we present in this paper focuses on furthering the understanding of
the physical requirements of disk
formation. The main question addressed by our feedback study  can be
phrased as follows:
Which of the many ingredients of the range of models summarized above are
necessary, which are sufficient and how exactly do they influence the galaxy
formation process alone and in combination? As in our study of the
angular momentum problem \citep[][Paper 1 hereafter]{Paper1}, we follow our strategy
of slowly increasing the complexity of the model. First using a preselected
dark matter halo with a mass similar to the halo of the Milky Way, with a high
spin parameter and a quiet merging history, usually thought to be the most
likely to host a disk, we study different methods of distributing feedback
energy. The best result becomes our standard model, which we then apply to a
set of halos with a range of different masses and a second set of Milky Way-type 
halos but with different merging histories. We also study a set of simulations examining
other relevant physical effects. We carefully analyze the resulting disk to
understand the influence each of these effects has alone, before we put
several physical effects together for our most realistic model of galaxy
formation. \par
The paper is organized as follows. In $\S$2 we describe the initial
conditions, the code and the analysis methods we use. The study of feedback
energy distribution methods is covered in $\S$3. $\S$4 contains the study of
halos with different assembly histories, $\S$5 the study of halos with
different masses and $\S$6 the application of additional physics besides the
standard feedback model. A detailed description of our most complex model can
be found in $\S$7 and the influence of the additional physics on the scaling
relations is discussed in $\S$8. We discuss our findings and conclude in $\S$9.

\section{Initial conditions, code and analysis methods}
\label{sec:IC_ana}
\subsection{Code}
The simulations are done with the N-body code \textsc{GADGET2}
\citep{springel05} using the Smoothed Particle Hydrodynamics (SPH) framework
(\citet{gm77}, \citet{lucy77}) for the gas and an implementation of radiative
cooling courtesy of Volker Springel and based on \citet{katz96}. For
simplicity, in the
standard cases we do not include an external UV background. Star formation is implemented
following \citet{Katz1992} and is already described in detail in Paper 1. It
is motivated by a Schmidt law \citep{schmidt59} giving the star formation rate as 
\begin{equation}
\frac{d\rho_{\star}}{dt} = c_{\star} \frac{\rho_{gas}}{t_{\star}}
\end{equation}
with $\rm{c_{\star}}$=0.1 and $\rm{t_{\star}}$=max($\rm{t_{dyn}}$,$\rm{t_{cool}}$),
and then the application of a stochastic approach. We assume that each gas
particles can spawn two generations of stars, so each star
particle has half the mass of a gas particle. Necessary conditions for star
formation are a critical density of $\rho_{crit}$=7$\times10^{-26}$g
cm$^{-3}$, a converging flow and a low temperature of T$<3\times10^{4}$ K.
\subsection{Initial conditions}
All of our halos are resimulations \citep[see][]{navarro94} from a large
cosmological box with 64 h$^{-1}$ Mpc$^3$ box size. We use the WMAP3
cosmology \citep{spergel07} with $\rm{H}_0$=73~$\rm{km}\,s^{-1}\,Mpc^{-1}$,
$\sigma_{8}$=0.75, n$_{rm{s}}$=0.9, $\Omega_{0}$=0.24 and
$\Omega_{\Lambda}$=0.76. The simulations start at z=50. Gas particles are
added on top of each dark matter particle assuming the cosmological baryon
density, $\Omega_{\rm{bar}}$=0.04. Our standard halo (MW$\_$mr) is a halo with a mass
similar to the Milky Way \citep[about $10^{12}\,\rm{h}^{-1}\,\Msun$][]{Smith2007,Xue2008}, with a fairly
quiet merging history and a comparatively high spin parameter
$\lambda$=0.054. Our standard resolution corresponds to 1024$^3$ effective particles
within the high resolution region surrounding our target halo. This translates
to particle masses of $2.8\times10^6\,\rm{h}^{-1}\,\Msun$ for gas and
$1.34\times10^7\,\rm{h}^{-1}\,\Msun$ for dark matter particles, and we use a
softening of 1.5 and 2 h$^{-1}$ kpc respectively. Additionally, we selected three halos comparable in mass to the
standard halo, but with very different merging histories, to study the
influence of the merging history of the host halo on the formation of the
disk. Lastly, we selected six halos with smaller masses starting at
$\approx10^{11}\,\Msun$ and one halo with a larger mass of
$\approx2\times10^{12}\,\Msun$ to sample the disk galaxy scaling relations
over a larger mass range. All halos were selected to have no object of equal
or larger mass close by. All the lower mass halos and also the standard halo
have been simulated in high resolution with 2048$^3$ effective particles in
the resimulation region and particle masses and gravitational softening
parameters of $3.54\times10^5\,\rm{h}^{-1}\,\Msun$,
$1.7\times10^6\,\rm{h}^{-1}\,\Msun$ and 0.75 and 1 h$^{-1}$ kpc,
respectively. Table~\ref{tab:halos} shows an overview of characteristic
parameters of the halos we simulate. \par
\begin{table*}
\begin{tabular}{rrrrrrrr}
\hline
\hline
halo&M$_{\rm{vir}}$&R$_{\rm{vir}}$&V$_{\rm{vir}}$&$\lambda$&z$_{\rm{lmm}}$&effective&N$_{\rm{vir}}$\\
&10$^{12}\,\Msun$&kpc&km s$^{-1}$&&(1:3)&resolution\\
\hline
\hline
MW$\_$mr&1.26&296&146.17&0.054&3.63&1024$^3$&$1.6\times10^5$\\
\hline
MW$\_$hr&1.23&290&143.22&0.058&3.63&2048$^3$&10$^6$\\
\hline
MW$\_$mh1$\_$mr&1.24&294&145.66&0.023&4.85&1024$^3$&$1.55\times10^5$\\
\hline
MW$\_$mh2$\_$mr&1.23&298&137.2&0.063&1.18&1024$^3$&$1.8\times10^5$\\
\hline
MW$\_$mh3$\_$mr&1.2&291&144.2&0.05&1.35&1024$^3$&$1.5\times10^5$\\
\hline
DM$\_$hr1&0.135&139&68.32&0.04&4.56&2048$^3$&10$^5$\\
\hline
DM$\_$hr2&0.252&173&85.11&0.029&3.02&2048$^3$&$2.42\times10^5$\\
\hline
DM$\_$hr3&0.358&193&95.34&0.034&3.44&2048$^3$&$3.13\times10^5$\\
\hline
DM$\_$hr4&0.493&214&105.5&0.019&6.09&2048$^3$&$4.2\times10^5$\\
\hline
DM$\_$hr5&0.594&227&112.38&0.016&3.44&2048$^3$&$5\times10^5$\\
\hline
DM$\_$hr6&0.703&241&119.36&0.026&3.83&2048$^3$&$6.1\times10^5$\\
\hline
DM$\_$mr7&2.36&364&180.37&0.018&4.556&1024$^3$&$2.85\times10^5$\\
\hline
\hline
\end{tabular}
\caption{Characteristic parameters for our halos run with the standard
  feedback model. MW labels our standard, Milky Way-type halo, MW$\_$mh halos with
  the same mass but varying merging histories (mh) and DM$\_$hr1 to 7 halos with different
  masses. lr, mr and hr indicate the effective resolution of the high
  resolution region shown in the second to last column. N$_{\rm{vir}}$ is the approximate
  total number of gas, star and dark matter particles in R$_{\rm{vir}}$ in the
  runs with the standard feedback model. This slightly varies for different
  feedback models, but the order of magnitude stays the same.}
\label{tab:halos}
\end{table*} 

\subsection{Analysis}
\begin{figure}
\begin{center}
\includegraphics[width=0.5\textwidth]{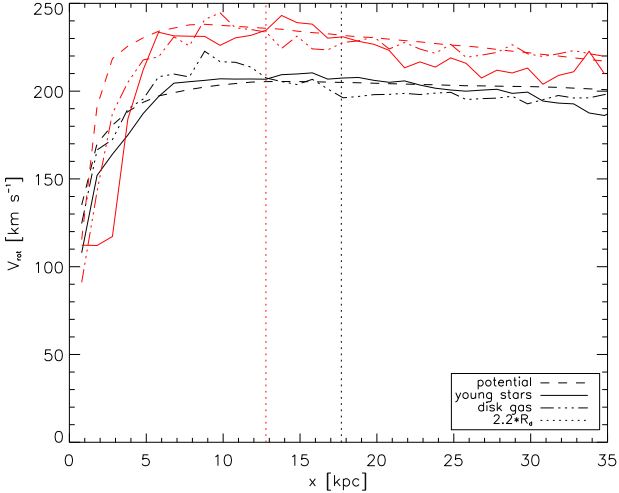}
\caption{Rotation curves for the standard feedback model in standard
  (red) and high resolution (black).The long dashed line dubbed
  'potential' is the circular velocity curve calculated from $V_{rot}=\sqrt{GM(<R)/R}$,
  the solid line traces the young stars in the rotationally supported disk,
  the dash-dotted line the cold disk gas. The vertical lines mark
  the location $2.2\rm{R_d}$, where the rotationally velocity is measured. }
\label{fig:vrot}
\end{center}
\end{figure}

We define our halos using R$_{\rm{vir}}$
via $\rho(R_{vir})=\Delta\rho_c$ with $\rho_c=3H_0^2/(8\pi G)$ and
$\Delta=18\pi^2+82x-39x^2$,
$x=\Omega_0(1+z)^3/(\Omega_0(1+z)^3+\Omega_{\Lambda})-1$. Characteristic
parameters for the galaxy are calculated using all particles within a sphere of
$\rm{6R_d}$. For the Milky Way with $\rm{R_d}\approx$3.5 kpc, this corresponds
to the disk radius. For our simulated galaxies with larger scale lengths,
$\rm{6R_d}$ corresponds approximately to the size of the Milky Way
stellar halo \citet{Schneider2006}. R$_{\rm{d}}$ is the disk scaling length. As gaseous disk we define the cold gas (with
$T<3\times10^4$K) in the center of the halo, which forms a clearly
distinguishable disk. In
most cases, there is only little hot gas surrounding the cold disk. We rotate the
galaxy so that the disk is in the x-y plane. Rotational velocities are measured
at $2.2\rm{R_d}$ using the rotation curve of the disk gas. This is close to
observational measures of rotational velocities which are often obtained using
HI observations \citep[for example][]{deBlok2008}.
An example for rotation curves with the standard halo and standard feedback model is shown in
Figure~\ref{fig:vrot} for the standard and the high resolution run. 
Our rotation curves are fairly flat, though particularly in the smaller halos,
and in the more bulge dominated models, they can be steeper than shown
here. However, the difference between the measured rotational velocity at the
peak and at the adopted radius, even for our steeper curves, is
never more than 10\% and therefore does not change our results
significantly. Particularly it cannot account for the offset of the zero-point
of the Tully-Fisher relation we see for our simulated galaxies.\par
The angular momentum of the galaxy is calculated for the gas and stars of the galaxy
as described in Paper~1, by summing over all particles, with respect to the
center of mass. This angular momentum is 
compared to the value expected from the properties of dark matter halos 
$\rm{j_{calc}}\approx1.3\times10^3(\rm{V_{rot}}/200)^2\,\rm{km}\,s^{-1}\,h^{-1}\,kpc$
as derived by \citet{NavarroMS2000}.\par
In order to study the importance of the disk in our galaxies, we also perform
a dynamical decomposition following \citet{abadi03b}. We compare the z
component of the angular momentum $\rm{j_z}$ with the angular momentum of the
corresponding circular orbit $\rm{j_{circ}(E)}$ for each star particle. The ratio
$\rm{\epsilon_j=j_z/j_{circ}(E)}$ describes the degree of rotational support of a
given stellar particle. A thin, rotationally supported disk has
$\rm{\epsilon_j}\sim1$, a spheroid has little net rotation due to equal numbers of
stars on co- and counterrotating orbits,
its distribution of $\rm{\epsilon_j}$ peaks at zero. All stars fitting in neither
category are classified as thick disk stars. They are not rotationally
supported at the same level as the thin disk, but rotate in the same manner, and spatially form a
thick disk. The thin disk is generally dominated by young stars, but can also
contain older stars. The
dynamical bulge-to-disk ratios quoted in the tables are ratios of the mass of
the bulge to the combined mass of thin and thick disk.

\subsection{Comparison with observations}
\label{subsec:howto}
In order to perform a realistic comparison with observational results, particularly
with respect to observed scaling relations like the Tully-Fisher relation, we
create and analyze mock observations following a method described by
\citet{arman}. Using population synthesis models by \citet{BC03} (BC) and the
software \textsc{SKYMAKER}\footnote{Bertin \& Fouqu\'e (2007),\\
  http$://$terapix.iap.fr$/$rubrique.php?id$\_$rubrique=221}, we mimic
observations performed by the Hubble Space Telescope. For this, the
metallicity of each star particle is needed. For simplicity, in our runs without explicit
metal enrichment, we assume solar metallicity for star particles younger than
10 Gyr, and Z=$10^{-4}$ for star particles older than 10 Gyr. We explore BC
models with Salpeter \citep{Salpeter} and Chabrier \citep{Chabrier} initial
mass functions (IMF). The main difference between the two is a flatter (more physical)
behavior at the low mass end for the Chabrier IMF, resulting in mass-to-light
ratios being a factor of 1.5 smaller than for a Salpeter IMF. The images shown in this paper combine U, V and B band
``data''. We also create I band FITS files centered on the galaxy which are used to
compute the surface brightness profile using the \textsc{ESO-MIDAS} package\footnote{www.eso.org/projects/esomidas}, which is then fitted with \textsc{GALFIT}
\citep{galfit}. To obtain the galaxy's total magnitude as well as a
photometric decomposition we fit a two component bulge+disk model to the whole
profile. To find the disk scale length, we fit a pure exponential model to the
outer parts. The rotational velocity is then measured with the rotation curve
for the cold gas at $2.2\rm{R_I}$. The I band luminosity $\rm{L_I}$
is calculated from the magnitude following \citet{courteau07} as
$\rm{L_I}$=10$^{-0.4(\rm{M_I}-4.19)}$.\par
Observers measure the angular momentum of a galaxy from the rotation curve and
the scale radius as
$j_{\rm{obs}}=2R_{\rm{d}}V_{\rm{rot}}(2.2R_{\rm{d}})$. This relation, also
known as the angular momentum-size relation, is
inferred for an exponential disk in an isothermal halo \citep{Fall1980}. For our simulated
galaxies, besides the real angular momentum content we also compute the
angular momentum in this manner, using the I-band scale length and rotational
velocity as described for the Tully-Fisher relation. This "exponential disk
estimator" is then used to
compare to the actual angular momentum to investigate possible systematics.

\section{Supernova energy distribution methods}
In cosmological simulations, a star particle basically corresponds to a single stellar
population (SSP). The energy from
supernova explosions is therefore summed over the whole SSP and then
distributed over the surrounding gas particles. As a first step in our
investigation of supernova feedback, we study three different ways of
distribution.  
\subsection{Implementation}
We assume a Miller-Scalo initial
mass function for the stellar population in one of our star particles
(\cite{millerscalo}) with a lower cutoff of 0.1 M$_{\odot}$ and an upper
cutoff of 100 M$_{\odot}$: 
\begin{equation}
\xi(M)= M_{\star}A \left\{\begin{array}{cl} M^{-1.25} & 0.1 < M < 1 \Msun\\
    M^{-2} & 1 < M < 2 \Msun\\ 2^{0.3} M^{-2.3} & 2 < M < 10 \Msun\\ 10 \
    2^{0.3} M^{-3.3} & 10 < M < 100 \Msun \end{array}\right.
\end{equation}
where A=0.284350751.
Stars with masses between 8 and 40 M$_{\odot}$
explode as type II supernova, each explosion yielding an energy of 10$^{51}$
ergs. This results in an energy of 1.21$\times10^{49}$ erg per solar mass
formed. This energy is smoothed over the neighboring gas particles using the
SPH smoothing kernel, so each neighbor gets an energy of 
\begin{equation}
\Delta E_{SN,i}=E_{SN,tot} \frac{W(| \overrightarrow{r_{i}}-\overrightarrow{r_{\star}}|,h_{\star})M_{i}}{\rho_{\star}}
\end{equation}
$h_{\star}$ is the smoothing length and regulates the number of gas particles
receiving feedback energy. For this we try two different approaches. Initially,
we use a fixed radius for the smoothing sphere of 1.37 h$^{-1}$ kpc, which
ensures a distribution of energy consistent with resolution. However, in this
scheme, the impact of the feedback energy is highly dependent on the local
density, with the same amount of energy distributed to large or small numbers
of particles. The structure of the disks is
improved when the size of the smoothing region scales with the local density
instead, as does the smoothing length for SPH calculations in Gadget. With
this scheme, the mass-weighted number of neighbors receiving feedback energy
stays constant.\par  
\begin{table*}
\begin{tabular}{rrrrrrrrrrrrr}
\hline
\hline
run&M$_{\rm{bar}}^1$&M$_{\rm{bar}}^2$&f$_{\rm{bar}}$&f$_{\rm{cold}}$&SFR$^3$&R$^4_{\rm{d,I}}$&V$^5_{\rm{rot}}$&L$^6_{\rm{I}}$&(M/L)$_{\star}$&B/D&B/D&j$_{\rm{bar}}/$\\
&&&&&&&&&&dyn.&photo.&j$_{\rm{calc}}$\\
\hline
\hline
MW$\_$mr$\_$nf&14.4&14.4&0.163&0.056&1.87&6.09&224.8&5.65&2.41&0.91&0.92&0.55\\
\hline
MW$\_$mr$\_$i&7.49&8.1&0.126&0.173&0.89&7.9&207.71&2.63&2.36&0.31&2.61&0.95\\
\hline
MW$\_$mr$\_$inc&6.17&7.53&0.141&0.127&2.52&5.85&187.46&3.27&1.62&0.57&1.84&0.9\\
\hline
MW$\_$mr$\_$h&10.7&11.0&0.136&0.119&1.12&6.18&239.37&3.82&2.52&0.75&5.06&0.48\\
\hline
MW$\_$mr$\_$hnc&7.07&8.91&0.146&0.079&3.86&5.36&186.67&4.23&1.48&0.42&0.14&0.86\\
\hline
MW$\_$mr$\_$e&10.05&10.55&0.134&0.119&1.51&7.81&221.06&3.66&2.47&0.5&2.48&0.71\\
\hline
MW$\_$mr$\_$enc&8.05&8.7&0.136&0.094&1.31&7.03&221.2&3.3&2.21&0.35&0.59&0.7\\
\hline
MW$\_$mr$\_$encv&10.4&10.8&0.143&0.104&2.55&6.01&243.1&4.63&2.08&0.43&0.27&0.62\\
\hline
MW$\_$hr$\_$encv&7.2&7.82&0.113&0.11&1.38&7.89&207.47&3.13&2.06&1.02&1.49&0.9\\
\hline
\hline
\end{tabular}
\caption{Characteristic parameters for the galaxies with different
  distribution methods for the feedback energy. V$_{rot}$ is measured for young stars at 2.2R$_{\rm{d,I}}$. The luminosity
  and mass-to-light ratio are based on a Chabrier IMF for the edge-on projection. The last
  column measures the angular momentum content and contains the ratio of the
  total angular momentum of stars and gas in the galaxy and the expected value
  based on the relation
  $j_{calc}\approx1.3\times10^3(V_{rot}/200)^2\,\rm{km}\,s^{-1}\,h^{-1}\,kpc$
  from \citet{NavarroMS2000}. \newline
  $^1$ galaxy mass including cold gas only, in
  10$^{10}\,\Msun$; $^2$ galaxy mass including all gas, in 10$^{10}\,\Msun$;
  $^3$ in $\Msun~\rm{yr}^{-1}$; $^4$ in kpc; $^5$ in km~s$^{-1}$; $^6$ in
  10$^{10}\Lsun$}
\label{tab:expdist}
\end{table*} 

\begin{figure}
\begin{center}
\includegraphics[width=0.5\textwidth]{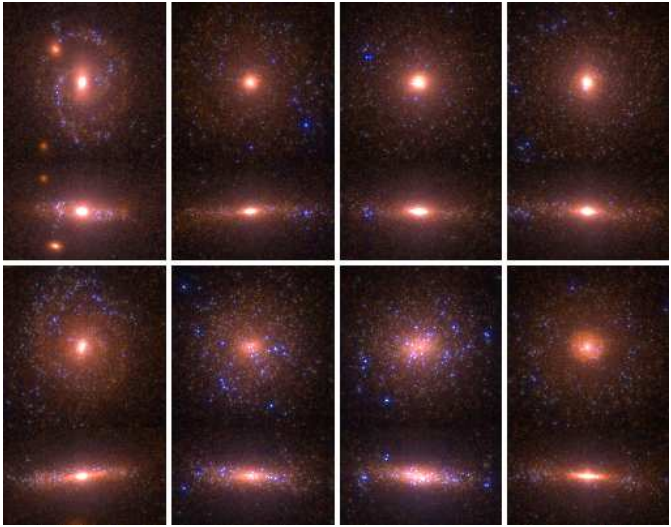}
 \caption{Mock observations in
  face on and edge on projections of our simulated galaxies with different
  methods of distributing feedback energy as described in the text and
  in Table~\ref{tab:expdist}. Each box has a side length of 60 kpc, and those
  with the edge-on projection a height of 30 kpc. In the top row we have the
  model without feedback, the instantaneous model, the heat-rate model and the
  exponential model. The models in the bottom row correspond to those on the
  top, but with cooling locally turned off. The bottom left panel is our
  standard model (exponential with local cooling turnoff and variable
  smoothing length).}
\label{fig:expdistsky}
\end{center}
\end{figure}
In the following, we briefly describe the three methods of supernova energy distribution. The
\emph{instantaneous approach} is the simplest and also most drastic
method. The total supernova energy is distributed to the neighboring gas
particles immediately  after formation of the star particle. \par
Dumping the full amount of supernova energy in a single timestep is a very
crude way of mimicking a real supernova remnant which expands and interacts
with the surrounding ISM over an extended period of time, typically 20-30 Myr. The following
two methods therefore distribute the energy more smoothly over time. In the \emph{heat-rate
  approach} we distribute the energy in equal portions over a total time of
t$_{\rm{SN}}$=20 Myr by 
\begin{equation}
\Delta E = E_{SN} \frac{\Delta t}{t_{SN}},
\end{equation}
where $\Delta t$ is the current timestep of the star particle.
A second possibility for a slow energy output is the \emph{exponential
  approach}, an exponential feedback model using
\begin{equation}
\Delta E = E_{SN} \frac{t-t_{\star}}{t_{SN}} e^{- \frac{t-t_{\star}}{t_{SN}}} \frac{\Delta t}{t_{SN}}.
\end{equation}
Again, t$_{\rm{SN}}$=20 Myr and t$_{\star}$ is the formation time of the stellar
particle. This model mimics effects owing to the formation time and lifetimes
of stars with different masses.\par
We also test variations of all of these methods where radiative cooling is turned off
temporarily in the gas particles receiving supernova energy \citep{Gerritsen1997}. This helps to prevent
immediate re-radiation of the thermal energy due to the high density of the
gas surrounding the star
formation sites and the inability to resolve the multiphase structure of the
interstellar medium. We choose a turnoff time period of
20 Myr, based on typical cooling times for supernova blastwaves. With this we
attempt to suppress the strong early star formation more efficiently.
\subsection{Results}
\begin{figure}
\begin{center}
\includegraphics[width=0.5\textwidth]{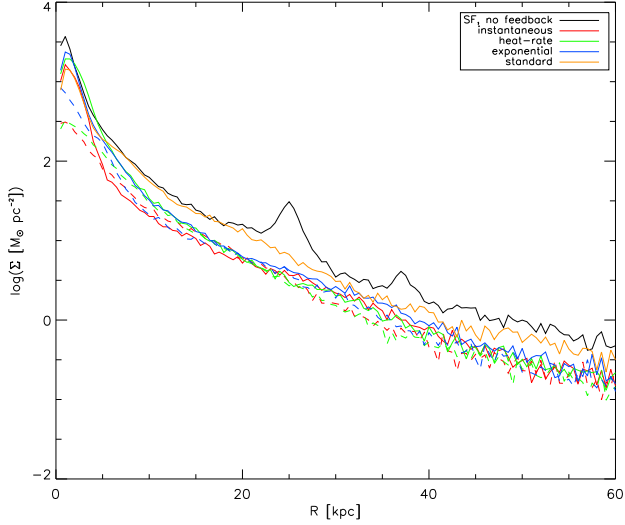}
\caption{Stellar surface density profiles for the different
  feedback models. The different colors indicate the different models as
  shown in the legend. The dashed lines show the corresponding model with
  local cooling turned off. The bump in the model without feedback and in the
  exponential model at about 25 kpc radius corresponds to an incoming
  satellite galaxy also clearly visible in Figure~\ref{fig:expdistsky}.}
\label{fig:surface_den}
\end{center}
\end{figure}
\begin{figure}
\begin{center}
\includegraphics[width=0.5\textwidth,angle=90]{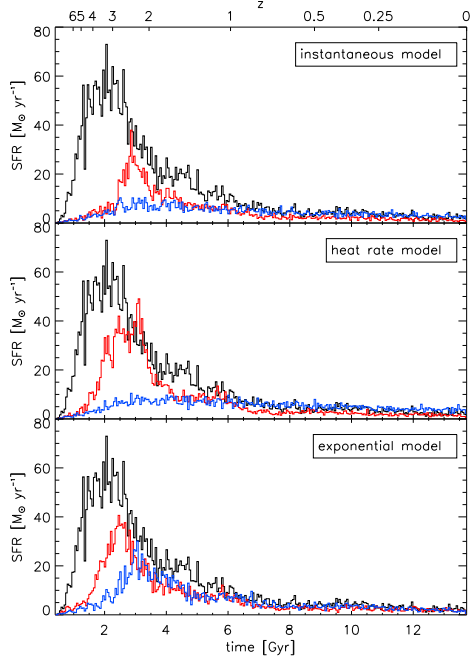}
\caption{Comparing the star formation histories for the different feedback
  energy distribution methods in comparison to the run without feedback (black
  line). The red line shows the standard case, the blue line the case with
  locally turned off cooling (``nc'').}
\label{fig:expdist_sfr}
\end{center}
\end{figure}
\begin{figure}
\begin{center}
\includegraphics[width=0.5\textwidth]{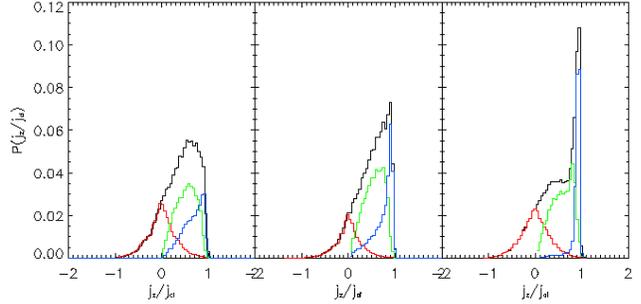}
\caption{Dynamical decomposition of the stellar component for the heat
  rate model (left panel), the heat-rate model with variable smoothing length
  (middle panel) and the exponential model with exponential smoothing length
  (right panel), all with local cooling turnoff. The black line represents all
  stars, the blue line thin disk stars, the green line thick disk stars and
  the red line the spheroidal component.}
\label{fig:decomp1}
\end{center}
\end{figure}
\begin{figure}
\begin{center}
\includegraphics[width=0.5\textwidth]{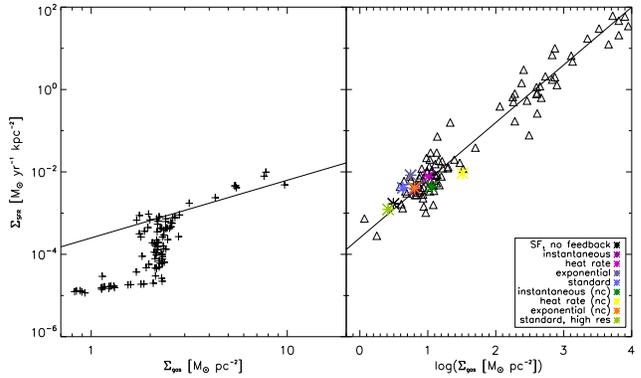}
\caption{Left panel: Agreement of the galaxy formed in the
  standard high resolution model with the Kennicutt law, locally given by
  equation~\ref{eq:kennicutt} (solid line). In the outer region, 
  the star formation rate drops since the disk is no longer Toomre
  instable. Right panel: Agreement of all our galaxies with different feedback
  distribution methods with the observed global Kennicutt law. The triangles are
  observational data points from \citet{Kennicutt1998}. Our model galaxies
  populate the lower surface density region in agreement with the regular
  galaxies, while the observed galaxies with high surface densities are
  starburst galaxies.}
\label{fig:kennicutt}
\end{center}
\end{figure}
Table~\ref{tab:expdist} and Figure~\ref{fig:expdistsky} give an impression of
the results for the different energy distribution methods. Our standard halo
MW$\_$mr has been used as initial conditions for all of these runs. The abbreviations i
(=instantaneous), h (=heat-rate) and e (=exponential) indicate the type of
feedback, nc stands for the local cooling turnoff, and nf for no feedback which
is a run with only star formation which is included for comparison. All runs
are employing fixed smoothing length except one, MW$\_$mr$\_$encv, with variable
smoothing length for feedback energy. Variable in this case means that the
smoothing length is calculated following the smoothing length for the SPH
density calculation. It is scaled according to the mass-weighted number of
neighbors which is kept constant to a value of 40. We experimented with the variable smoothing length also in
the other cases but could not see substantial improvement.
The first obvious result from Figure~\ref{fig:expdistsky} is the importance of
turning off the cooling locally in order to form a young, bright
disk (bottom row vs top row). This is also clear from the stellar 
surface density profiles shown in Figure~\ref{fig:surface_den}. While all runs
with feedback are quite similar outside of about 8 kpc, the galaxies
with the cooling turn-off have a much flatter slope in the inner regions,
corresponding to a smaller bulge. In the simulation without any feedback, the bulge is most
dominant, as is expected. The two prominent spiral arms seen in the image in
the top left panel of Figure~\ref{fig:expdistsky} are due to ongoing
mergers of smaller satellites (one is also visible in the image) inducing new star formation.
Nevertheless the galaxy is very massive, slowly rotating and clearly
spheroidal with a very low angular momentum. The clearly visible incoming
satellite can be seen as a bump in the surface density profile. The bulge is
also important in runs with feedback but without the local cooling turn-off.\par
Figure~\ref{fig:expdist_sfr} shows the star formation histories for all cases
in comparison with the no-feedback run. We see that the instantaneous model even without
the local cooling turn-off suppresses early star formation most
efficiently due to the instantaneous energy input. The star formation history
for the two models with smooth energy input are very similar, except when
cooling is turned off locally, which has less influence in the exponential
model than in the heat-rate model. However, in the basic instantaneous model, star formation
is suppressed but not self-regulated which is why we do not see a young stellar
disk. The feedback actually becomes too strong. When we turn off cooling
locally, a significant amount of gas stays hot for a longer amount of time,
but also close to the disk, so when it cools
again it can form stars there and we find a young but rather thick and
unstructured stellar disk. In the exponential model, the suppression is less
efficient and the young disk less prominent.\par
A more prominent spiral structure is achieved when a
variable smoothing length is used, since in that case the code can better adjust to local
density structures. For the exponential model with cooling turnoff this is
shown in the bottom left image of Figure~\ref{fig:expdistsky}. Our two best cases in comparison to the Milky
Way are MW$\_$mr$\_$hnc and MW$\_$mr$\_$encv, i.e. the heat-rate and the 
exponential model with local cooling turnoff and, for the latter, with a
variable smoothing length. MW$\_$mr$\_$hnc has an almost
perfect exponential surface brightness profile and is almost not angular
momentum deficient. Nevertheless, as shown in Figure~\ref{fig:decomp1}, the dynamic
decomposition shows it to be rather a thick disk instead of a thin disk and a
bulge, and the structure of the disk is very clumpy. This does not improve
much when a variable smoothing length is used in this case.
We therefore select
MW$\_$mr$\_$encv as our standard model for the rest of our
work. It fits the Kennicutt relation, given by
\begin{equation}
\Sigma_{\rm{SFR}}=(1.5\pm0.7)\times10^{-4}\left(\frac{\Sigma_{\rm{gas}}}{\Msun\,\rm{pc}^{-2}}\right)^{1.4\pm0.15}\frac{\Msun}{\rm{yr}\,kpc^2},
\label{eq:kennicutt}
\end{equation}
well as shown in the left panel of
Figure~\ref{fig:kennicutt}, though the outer regions are no longer Toomre
instable and the star formation density drops. In general, all of our models
fall well within the scatter of the observed global Kennicutt relation in the
region of regular galaxies, as
shown in the right panel of Figure~\ref{fig:kennicutt}. \par
We test the standard model with our higher resolution halo, as run 
MW$\_$hr$\_$encv, with about $10^6$ gas, star and dark matter particles within
the virial radius. Our feedback method does depend somewhat on resolution with
the main effect being a more efficient suppression
of star formation which reduces the total galaxy mass as measured in a radius
of 6R$_{\rm{d}}$ by about 30\%. The high resolution galaxy has basically no angular momentum deficiency. Even though the
scale length in the high resolution case is a little larger, the surface
brightness profiles are very similar. Overall the higher resolution is
slightly different in some aspects but the model is reasonably robust and the
changes at higher resolution improve the disk rather than having a negative impact.

\section{Influence of different assembly histories}
\begin{figure}
\centering
\subfigure[]
{
  \includegraphics[width=0.5\textwidth]{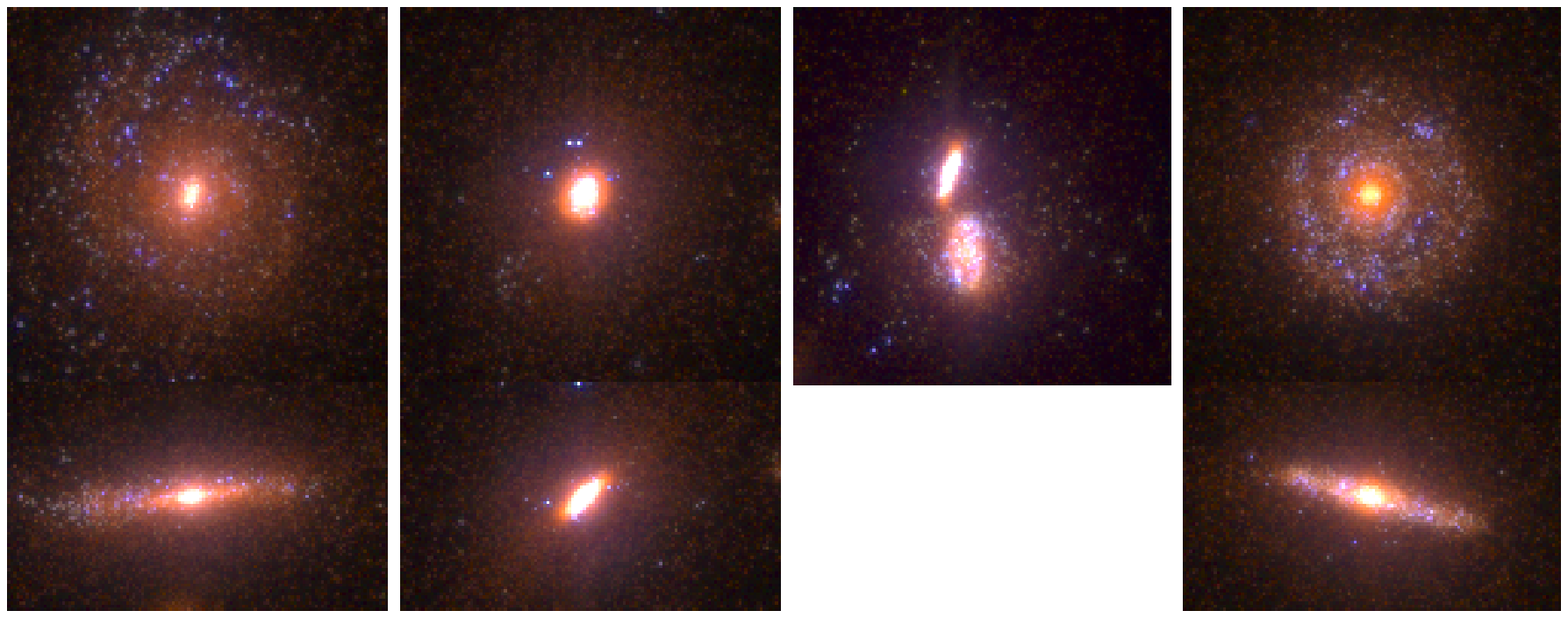}
  \label{fig:mergers1}
}
\subfigure[]
{
  \includegraphics[width=0.5\textwidth]{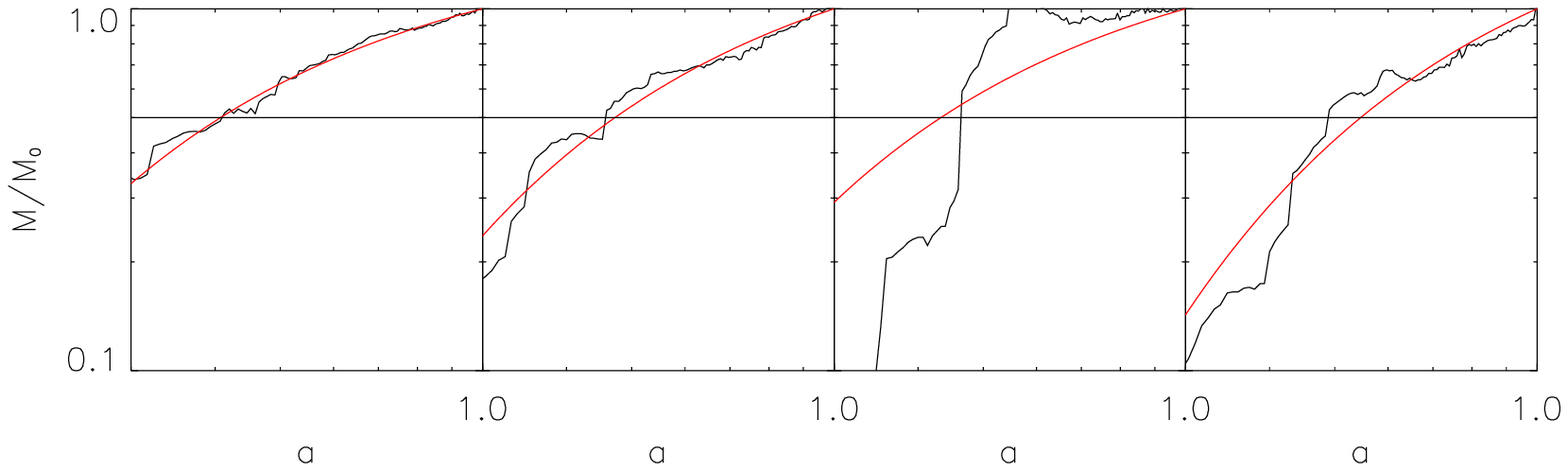}
  \label{fig:mergers2}
}
  \caption{Panel (a): Mock observations
  of the Milky Way type galaxies with different merging histories, all run
  with the standard feedback model. For comparison the standard
  halo is shown on the left, followed from left to right by MW$\_$mh1$\_$mr, MW$\_$mh2$\_$mr
  and MW$\_$mh3$\_$mr. There is no edge-on projection for MW$\_$mh2$\_$mr,
  since this is still in the process of merging. Panel (b): The halo mass
  growth in dependence of the scale parameter a, fitted with the model given by Equation~\ref{eq:wechsler} to assess how typical the halo is. }
\label{fig:mergers}
\end{figure}
\begin{figure}
\begin{center}
\includegraphics[width=0.5\textwidth]{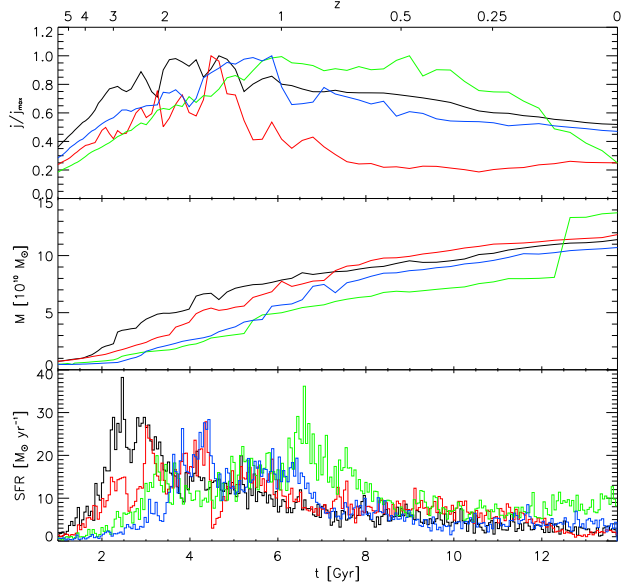}
\caption{The angular momentum evolution (top panel) and  mass
  growth (middle panel) for the combined stellar and gaseous components,
  and the star formation history (bottom panel) of the galaxies in halos
  with different assembly histories. The lines (black, red, green and
  blue) correspond to the halos as shown in
  Figure~\ref{fig:mergers1} from left to right, with the black line being our
  standard halo.}
\label{fig:merger_assembly}
\end{center}
\end{figure}
Owing to the persistent problem of forming realistic disk galaxies in
cosmological simulations over the past two decades, it became custom to choose a halo with what is
thought to be favorable conditions for the formation of a disk. This
comprises particularly a high spin parameter, resulting in disks with large
scale lengths, and no late large mergers which
could destroy the disk. The goal is to succeed in simulating at least one good
disk as a basis for further work. However, it has been shown recently by \citet{scannapieco09}
that disks can form in a whole range of halos with high or low spin parameters
leading these authors to question the previously assumed correlation between formation
history and morphology of a galaxy. \citet{governato08} showed that an
extended disk can form in a simulation even after larger mergers via
re-growth, if the progenitors are gas-rich. 
We therefore also apply our standard feedback model to three other halos with a similar Milky
Way-type mass as the standard halo, but with different merging histories. The only
criterion in picking these halos was their mass and a distance of at least 2
Mpc h$^{-1}$ to the next object of equal or higher mass in order to avoid
direct encounters at z=0. In
Table~\ref{tab:halos} these are the runs denoted by MW$\_$mh1$\_$mr,
MW$\_$mh2$\_$mr and MW$\_$mh3$\_$mr. The assembly histories
and mock observations are shown in Figure~\ref{fig:mergers}. To characterize,
how typical the chosen halos are, we fit the mass growth curve using
\begin{equation}
M(a) = M_0e^{-\alpha z}, a=(1+z)^{-1}
\label{eq:wechsler}
\end{equation}
from \citet{wechsler02} ($\alpha$ is a free parameter characterizing a
characteristic epoch of formation). For mass accretion histories without large mergers,
this provides a good description. This is the case for our standard halo, for
MW$\_$mh1$\_$mr and also, with some limitation,
MW$\_$mh3$\_$mr, as can be seen in Figure~\ref{fig:mergers2}. \par
Figure~\ref{fig:merger_assembly} shows the angular momentum, mass growth and
star formation history of
the galaxies compared to our standard halo. MW$\_$mh2$\_$mr is a quite extreme
case. Two large baryonic progenitors merge at
z$\approx$0.1, and as is obvious in the image in Figure~\ref{fig:mergers1}, this
process is not finished at z=0. The plot of the mass growth of the halo shows
only a very slight increase at these late times, after a very large (1.26:1)
merger at around z=1. The merger coincides with a large starburst.\par
Comparing the assembly histories of MW$\_$mh1$\_$mr and MW$\_$mh3$\_$mr, we
would have expected MW$\_$mh1$\_$mr to be more likely to have a disk, due to
the large mass increase at z=1 for MW$\_$mh3$\_$mr (where the mass is more than
doubled; this is preceded by a 3.14:1 dark matter merger). However, as
can be seen from the angular momentum evolution in
Figure~\ref{fig:merger_assembly}, the assembly is more chaotic for
MW$\_$mh1$\_$mr with a very large loss of angular momentum at
z$\approx$1.3 coinciding with a drop in star formation. After that, even though the disk still grows in mass, it does
not in size. At z=1 it already resembles very closely its final state. It is a low angular momentum system with $\lambda=0.023$,
compared to $\lambda=0.05$ for MW$\_$mh3$\_$mr. The latter, despite its 
later merger, looses significantly less angular momentum in its
merger. The angular momentum evolution and mass growth are both 
quite comparable to our standard model except for the delayed 
assembly. While in radius it is comparable to MW$\_$mh1$\_$mr at z$\approx$1,
its disk can grow due to higher angular momentum gas still entering the system.
Also, star formation sets in much later for this halo, peaking at
z$\sim$1.5 instead of 3 as in the standard model, due to the overall later
assembly (see Figure~\ref{fig:merger_assembly}). This helps in keeping the bulge
small. Star formation is generally slightly higher
for MW$\_$mh1$\_$mr (with $10\Msun\,\rm{yr}^{-1}$ from z$\sim$1 to z$\sim$0.25),
leading to a low final gas fraction of 0.015 compared to 0.084 for
MW$\_$mh3$\_$mr, and overall to a much more spheroidal shape. \par
We can conclude that a quiet and early assembly history of the dark
matter halo does not necessarily lead to a disk, confirming the conclusions by \citet{scannapieco09}. Not only the time of
the last major merger, but rather the onset of star formation and the impact particularly in loss of angular
momentum seems to mainly influence the disk. This could be related to the
geometry of the merging disks as suggested by \citet{scannapieco09}. The quick
growth of the disk in our MW$\_$mh3$\_$mr halo is very similar to what has
been reported by \citet{governato08}.

\section{The standard model over a range of masses}
\begin{figure}
\begin{center}
\includegraphics[width=0.5\textwidth]{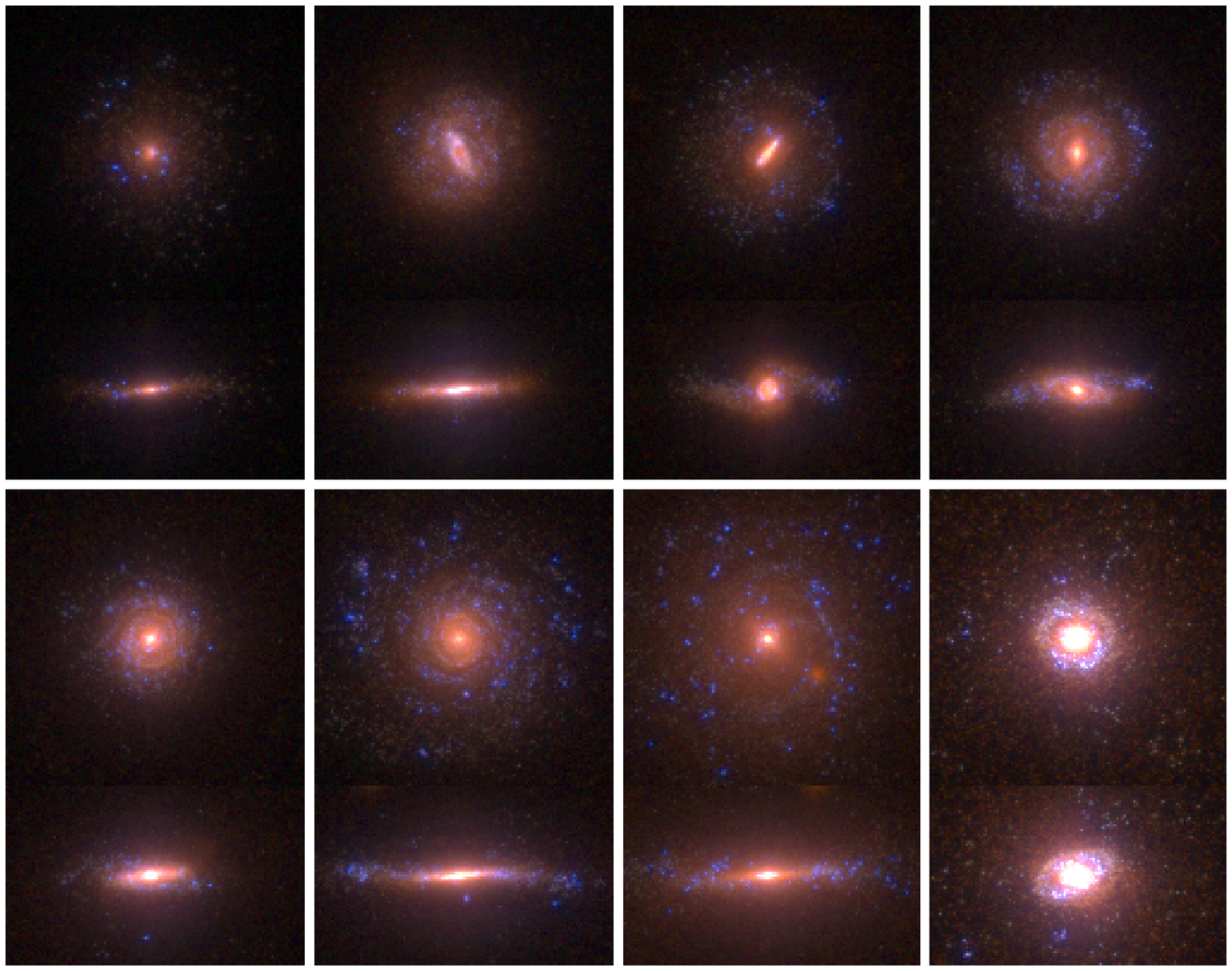}
\caption{Mock observations of runs of halos spanning a mass range of an order of
  magnitude, from $1.13\times10^{11}$ to $1.94\times10^{12}\,\Msun$ in dark
  matter halo virial mass. The top row shows runs  DM$\_$hr1, DM$\_$hr2,
  DM$\_$hr3 and DM$\_$hr4, the bottom row has DM$\_$hr5, DM$\_$hr6,
  MW$\_$hr (the standard run in high resolution) and DM$\_$mr7, the only one
  in this set with the medium resolution of 1024$^3$ effective resolution.}
\label{fig:diff_masses}
\end{center}
\end{figure}

To test the performance of our standard model with respect to the observed
scaling relations of disk galaxies, we perform runs using halos spanning a
mass range from $1.13\times10^{11}$ to $1.94\times10^{12}\,\Msun$ in dark
matter halo virial mass. For all except the largest mass halo, we have
performed the runs with an effective resolution of $2048^3$ ("hr" runs in our
nomenclature). The halos were selected to have no late major mergers as well
as relatively high spin parameters. The latter did not necessarily impact the
disk. The galaxies forming in the halos DM$\_$hr3 and DM$\_$hr7 are not very
disk-like. While the latter indeed has a rather low spin parameter with
$\lambda=0.018$, the spin parameter of the former is $\lambda=0.034$, quite
close to the mean value of halos. All of the chosen halos have a fairly quiet
assembly history. The gallery of mock observations for
these runs is shown in Figure~\ref{fig:diff_masses}, with increasing mass from
top left to bottom right and the standard halo
in high resolution being the third image in the bottom
row. Table~\ref{tab:massrange} contains the characteristic parameters. \par
In our sample we have two very good cases with a strong rotationally supported
young thin disk: DM$\_$hr2 ($\rm{M_{halo}}=2.52\times10^{11}\,\Msun$) and DM$\_$hr6
($\rm{M_{halo}}=7.03\times10^{11}\,\Msun$), but their masses are too low to resemble the Milky Way. While DM$\_$hr2 has a fairly constant
star formation history (SFH) of about $1.75\,\Msun\,\rm{yr}^{-1}$, the
other halos tend to have a
series of bursts showing the self-regulation of the feedback model. 
The surface brightness profiles of the rotationally supported disks are very
close to exponential, which can also be seen from their low photometric bulge-to-disk (B/D) ratios showing a
bright and prominent disk. For DM$\_$hr6, this is 0.53, the lowest of the set.
Dynamically, the bulge tends to be stronger except for DM$\_$hr2, where
(B/D)$_{\rm{dyn}}$=0.27 only. Our highest mass halo does not
host a disk galaxy, but a very luminous object. Possibly due to the lower
resolution it was simulated with, the disk never really grows to an
appreciable size. Therefore it is not expected to follow the scaling
relationships for disk galaxies, as will be confirmed in the following
section. This result is not surprising, since observationally, the most
luminous galaxies are indeed elliptical galaxies.

\begin{table*}
\begin{tabular}{rrrrrrrrrrrrr}
\hline
\hline
run&M$_{\rm{bar}}^1$&M$_{\rm{bar}}^2$&f$_{\rm{bar}}$&f$_{\rm{cold}}$&SFR&R$_{\rm{d,I}}$&V$_{\rm{rot}}$&L$_{\rm{I}}$&(M/L)$_{\star}$&B/D&B/D&j$_{\rm{bar}}/$\\
&&&&&&&&&&dyn.&photo.&j$_{\rm{calc}}$\\
\hline
\hline
DM$\_$hr1&0.95&0.97&0.11&0.326&0.3&3.26&105.44&0.39&1.93&2.23&0.25&0.83\\
\hline
DM$\_$hr2&2.44&2.49&0.134&0.135&1.75&3.81&136.92&1.5&1.41&0.27&0.69&0.9\\
\hline
DM$\_$h3&2.8&2.84&0.124&0.158&0.77&3.43&156.06&1.31&1.89&0.86&3.37&0.4\\
\hline
DM$\_$hr4&4.07&4.13&0.111&0.112&1.17&3.6&186.49&1.86&1.98&0.5&0.68&0.31\\
\hline
DM$\_$hr5&4.57&4.62&0.113&0.09&1.52&3.01&199.01&2.35&1.8&2.06&1.38&0.29\\
\hline
DM$\_$hr6&5.38&5.54&0.119&0.168&1.75&5.08&194.81&2.58&1.8&0.91&0.53&0.67\\
\hline
MW$\_$hr&7.2&7.82&0.113&0.11&1.38&7.89&207.47&3.13&2.06&1.02&1.49&0.9\\
\hline
DM$\_$mr7&1.75&1.8&0.137&0.048&6.7&3.48&319.64&8.73&1.91&0.42&0.56&0.09\\
\hline
\hline
\end{tabular}
\caption{Characteristic parameters for the resulting galaxies in the standard
  model applied to halos of increasing masses. \newline
  $^1$ galaxy mass including cold gas only, in
  10$^{10}\,\Msun$; $^2$ galaxy mass including all gas, in 10$^{10}\,\Msun$; $^3$ in $\Msun~\rm{yr}^{-1}$; $^4$ in kpc; $^5$ in km~s$^{-1}$; $^6$ in 10$^{10}\Lsun$}
\label{tab:massrange}
\end{table*} 

\subsection{Comparison with observed scaling relations}
\label{subsec:scale_rel}
The most important observed scaling relations with which one can test
simulations of disk galaxies are the
Tully-Fisher relation, the baryonic Tully-Fisher relation and relations with
respect to the angular momentum. We will discuss all of these in the following
sections.
\subsubsection{The angular momentum relations}
\begin{figure*}
\begin{center}
\includegraphics[width=\textwidth]{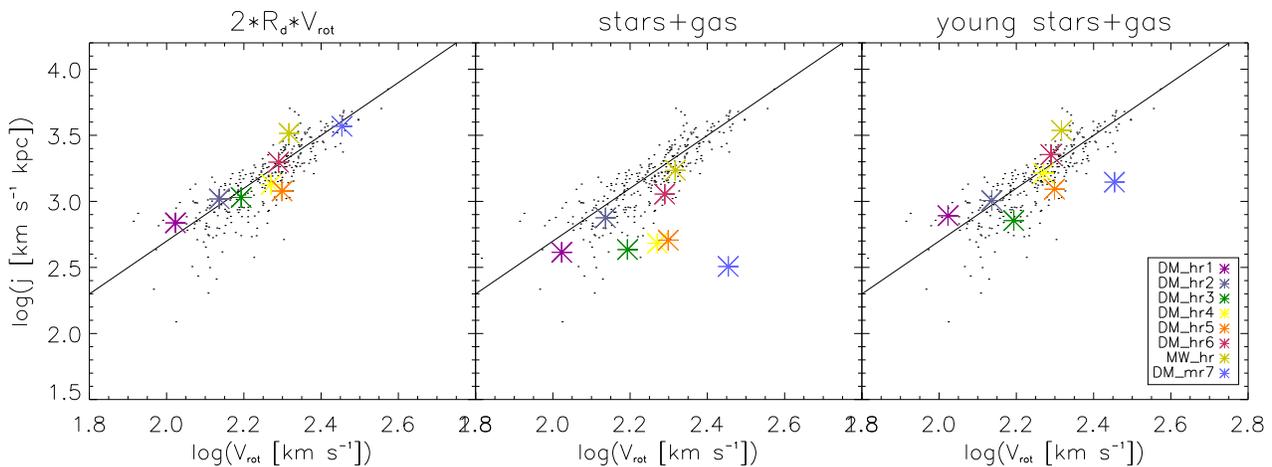}
\caption{Comparing the simulated galaxies to the
  observed angular momentum relation (dots), well fitted by the prediction from \citet{NavarroMS2000} (line). The
  left panel shows the angular momentum-size relation, the middle panel the
  true angular momentum of gas and stars in the simulated galaxies, and the
  right panel the angular momentum of the rotationally supported disk of young
  stars and gas.}
\label{fig:halo_am}
\end{center}
\end{figure*}
\begin{figure}
\begin{center}
\includegraphics[width=0.5\textwidth]{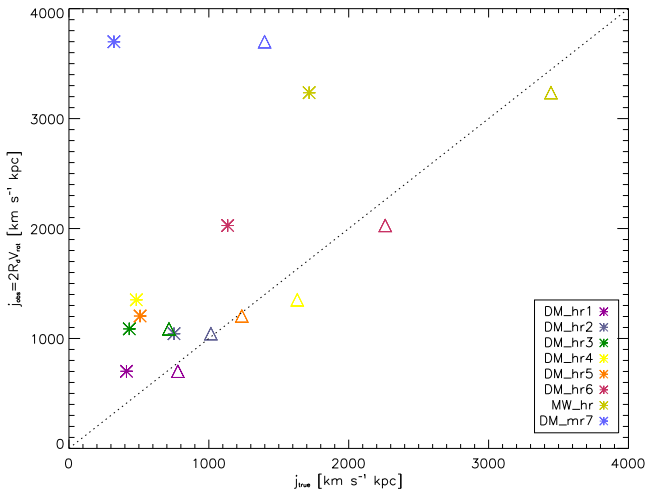}
\caption{The true, directly measured angular
  momentum of the simulated galaxies vs. the exponential disk estimator 
  $2R_dV_{rot}$. The dotted line indicates a direct
  correspondence. The star symbols are for the angular
  momentum of the whole galaxy, the triangles for the disk (gas and young
  stars) only.}
\label{fig:halo_am_true}
\end{center}
\end{figure}
For the angular momentum, we discuss two relations: the angular
momentum-size relation (as the "exponential disk estimator" from Section~\ref{subsec:howto}) and the relation of
the rotational velocity with the actual angular momentum of the galaxy. Our
methodology to measure the scaling length, rotational velocity, and angular
momentum is described in Section~\ref{sec:IC_ana}. We compare with observed
data from \citet{giovanelli}. Figure~\ref{fig:halo_am} shows the two
relations. The solid line is the expected relation (based on halo properties)
from \citet{NavarroMS2000}, 
$\rm{j_{calc}=1300(V_{rot}/200)^2\,km\,s^{-1}h^{-1}kpc}$. The left panel shows the
angular momentum-size relation. Since the rotation curves of disks are mostly
flat, this translates to a measure for the extend of the disk. For this we calculate j using
$\rm{j=2R_dV_{rot}}$ \citep{MMW1998}. The disk sizes of our simulated galaxies are
well within the observational scatter of the relation. The only
outlier is, as expected, DM$\_$hr7. Our disks therefore do not suffer from
being too compact and centrally concentrated, as in earlier
simulations. Looking at the middle panel, which
plots the actual angular momentum content of gas and stars in the galaxy, we
see that even though our galaxies tend to be slightly below the expected
relation, again they fall within the observed scatter. DM$\_$hr1, DM$\_$hr2,
MW$\_$hr and also DM$\_$hr6 have almost no angular momentum deficiency, while
DM$\_$hr5 is quite deficient due to a large amount of counterrotating gas. If we
only look at the actual disk consisting of cold gas and young stars, our data
scatter right around the expected relation. The difference between the angular
momentum as calculated from $\rm{j=2R_dV_{rot}}$, the exponential disk
estimator, and the actual angular momentum
content could be equivalent to a systematic difference between angular
momentum measured by observers and angular momentum derived from simulations. This
prompted us to compare these values directly, as shown in
Figure~\ref{fig:halo_am_true}. We indeed see a clear offset for the total
angular momentum content, while the angular momentum of the \emph{disk} (i.e. of disk
gas and young stars) is directly correlated with the 'observed' values. Two
conclusions arise from this. First, a fair comparison between observations and
simulations with respect to angular momentum is not straightforward when the
simulated galaxies have large bulge components. This is the case for those
of our galaxies which show the largest difference, DM$\_$hr3, DM$\_$hr4 and
DM$\_$hr5. Our standard feedback model is not able to produce a bulge-less
disk and only one clearly dominated by a disk, DM$\_$hr2 with B/D=0.27. The
second conclusion is that our actual disks composed of young stars and cold
gas have kinematic characteristics of real disks.

\subsubsection{The Tully-Fisher Relation}
\begin{figure}
\begin{center}
\includegraphics[width=0.5\textwidth]{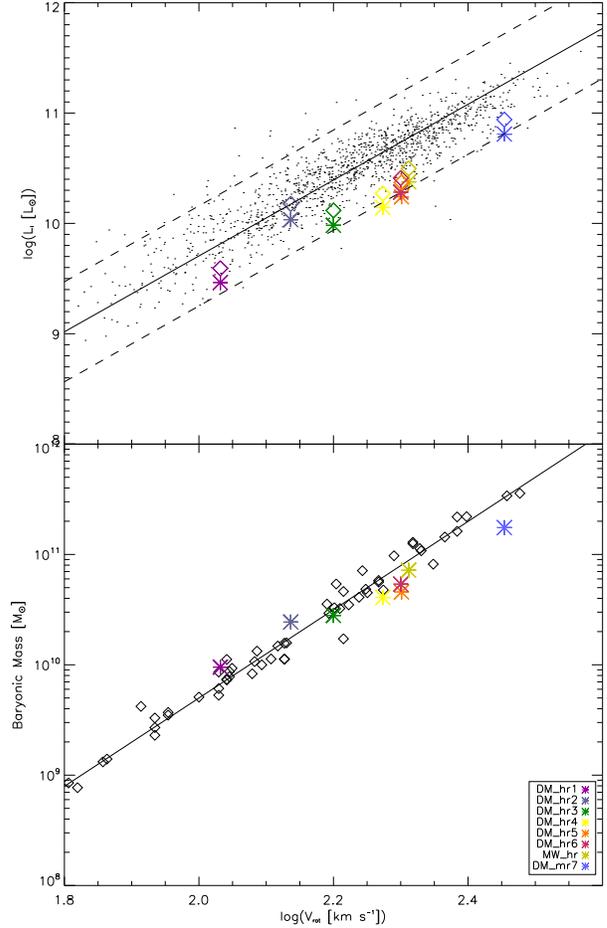}
\caption{The I-band (top) and baryonic (bottom) Tully-Fisher relations for our
  halos of different masses. The black points in the top panel are
  observations from the sample of \citet{courteau07}. The solid line
  is a fit, the dashed lines the 2$\sigma$ observed scatter. The stars are for
  a Salpeter IMF, the diamonds for a Chabrier IMF for our simulated galaxies. In the bottom
  panel, the black diamonds are observational data from \citet{mcgaugh2005}
  with a fit shown again by the solid line.}
\label{fig:halo_TF}
\end{center}
\end{figure}
The Tully-Fisher relation is one of the fundamental benchmarks for a
successful model of disk formation, and one of the biggest challenges of
simulations, since so far it remains difficult to reproduce the slope and the zero-point of the observed
relation. It relates the rotational velocity with the total luminosity of the galaxy and
is most often measured in the I or K-band since this is most representative of the
total stellar mass. We compare our results to observations
compiled by \citet{courteau07}. The relation is of type $L\propto V^{\alpha}$
and their fit to the $\log{\rm{V}}$ vs. $\log{\rm{L_I}}$ plot gives a slope of
$0.291\pm0.004$ and a zero point of $-0.835\pm0.039$ with a scatter
$\sigma_{\ln{V|L}}=0.132$. The top panel of Figure~\ref{fig:halo_TF} shows our
data overplotted on this relation. Our standard model misses the zero point of the relation by being too faint by
about a magnitude when a Salpeter IMF is used. The simulated galaxies barely fall within the
scatter. The only exception is the galaxy in halo DM$\_$hr2 which comfortably
lies within the scatter, though still below the best fit relation by half a
magnitude. A slight improvement can be achieved when using a Chabrier
IMF, shown by diamond symbols in Figure~\ref{fig:halo_TF}, but for most
galaxies the discrepancy is still large.\par 
The most fundamental relation to which the Tully-Fisher relation can be traced
is the baryonic Tully-Fisher relation \citep[see][]{mcgaugh2005}) relating
rotational velocity to total baryonic mass. We compare
to it in the bottom panel of Figure~\ref{fig:halo_TF}. \citet{mcgaugh2005}
finds as best fit to his observations the relation
$\rm{M_{bar}}=50V_{\rm{rot}}^4$. All simulated galaxies are well within the
observed scatter. We
seem to get a slightly shallower relation, though we do not really have enough
data points for a fit and the low mass halos also might be influenced by resolution
effects. With the exception of our expected outlier, DM$\_$mr7, we find the agreement to
be quite good.\par
From the results of the scaling relation comparisons we conclude that our
model is able to produce galaxies which are correct structurally, but still
bulge dominated and not luminous enough. Our I band stellar mass-to-light ratios of
M/L$\approx$2 (for a Chabrier IMF) are improved compared to earlier
simulations for example by \citet{NavarroMS2000} (who had M/L$\approx$2.5), but
are still too high compared to analytical expectations based on \citet{MMW1998}
in combination with the observed Tully-Fisher relation which yield
M/L$\approx$1-1.5. We suspect this to be the main reason for the failure in
fitting the Tully-Fisher zero point, rather than an incorrect rotational velocity
due to peaked rotation curves. Another effect could be that the
concentration of the dark matter halo is too high resulting in too high rotational
velocities \citep{NavarroMS2000a}. 

\section{Including additional physical processes}
\label{sec:code_diffphys}
\begin{figure}
\begin{center}
\includegraphics[width=0.5\textwidth]{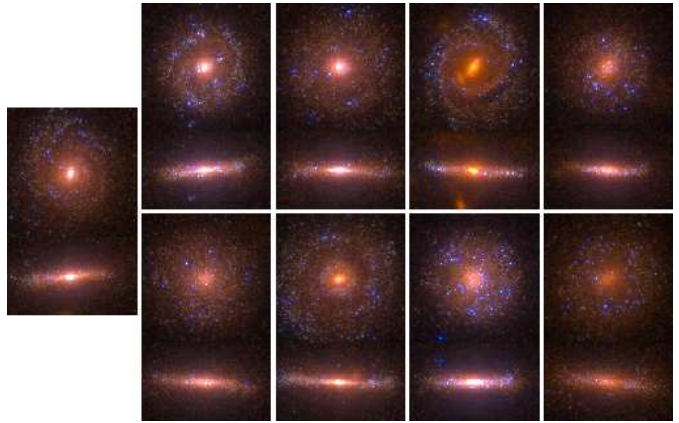}
\caption{Mock observations of the galaxies run with the standard feedback
  model and different additional physics effects. For comparison the standard
  model itself is shown on the left. The top row shows the runs with UV
  background, type Ia supernovae, metal-dependent cooling and the "all in"
  model with metal-dependent cooling. In the bottom row we show the runs with
  10\% kinetic feedback, mass return, the "all in" model and the blastwave model.}
\label{fig:physics_sky}
\end{center}
\end{figure}

The physical processes playing a role in the formation and evolution of a
galaxy are much more complex than just simple star formation and energy
feedback. In this section we describe a number of other relevant physical
effects we tested to study their influence on our standard model galaxy. An
overview of the characteristic parameters of the resulting galaxies can be
found in Table~\ref{tab:physics}, and Figure~\ref{fig:physics_sky} gives a
visual impression of the results.  \subsection{Implementation}
Before discussing the results we first briefly describe the implementation in each
case. The abbreviation in brackets is used later when referring to the simulations.
\begin{table*}
\begin{tabular}{rrrrrrrrrrrrr}
\hline
\hline
run&M$_{\rm{bar}}^1$&M$_{\rm{bar}}^2$&f$_{\rm{bar}}$&f$_{\rm{cold}}$&SFR&R$_{\rm{d,I}}$&V$_{\rm{rot}}$&L$\rm{_I}$&(M/L)$_{\star}$&B/D&B/D&j$_{\rm{bar}}/$\\
&&&&&&&&&&dyn.&photo.&j$_{\rm{calc}}$\\
\hline
\hline
standard&10.4&10.8&0.143&0.104&2.55&6.01&243.1&4.63&2.08&0.43&0.27&0.62\\
\hline
UV&9.64&11.0&0.165&0.056&4.07&5.74&218.02&5.22&1.74&0.41&1.79&0.65\\
\hline
kin&7.25&7.88&0.118&0.15&2.14&6.56&202.71&3.35&1.83&0.38&0.47&1.0\\
\hline
SNI&11.2&12.3&0.154&0.086&2.5&6.74&240.27&5.02&2.05&0.18&0.67&0.71\\
\hline
met&10.5&11.1&0.137&0.139&2.77&7.89&223.15&4.65&1.96&0.42&3.73&0.89\\
\hline
mSN&10.8&11.9&0.151&0.118&2.67&7.01&236.27&4.72&2.05&0.33&3.66&0.73\\
\hline
metc&15.2&15.3&0.164&0.105&1.64&7.07&242.34&5.41&2.57&0.33&2.75&0.64\\
\hline
bl&5.97&6.64&0.144&0.132&0.98&7.82&185.22&2.33&2.2&0.46&0.82&1.21\\
\hline
all in&9.59&11.1&0.157&0.154&5.7&5.67&225.15&5.6&1.51&0.26&1.17&0.78\\
\hline
all in, mc&7.5&7.52&0.09&0.162&1.85&5.08&204.28&3.24&1.97&0.44&1.01&0.72\\
\hline
\hline
\end{tabular}
\caption{Characteristic parameters for the resulting galaxies in the standard
  model with different additional physical effects. The standard model MW$\_$mr$\_$encv is
  shown again for comparison in the top row. \newline 
  $^1$ galaxy mass including cold gas only, in
  10$^{10}\,\Msun$; $^2$ galaxy mass including all gas, in 10$^{10}\,\Msun$; $^3$ in $\Msun~\rm{yr}^{-1}$; $^4$ in kpc; $^5$ in km~s$^{-1}$; $^6$ in 10$^{10}\Lsun$}
\label{tab:physics}
\end{table*} 
\begin{figure}
\begin{center}
\includegraphics[width=0.5\textwidth]{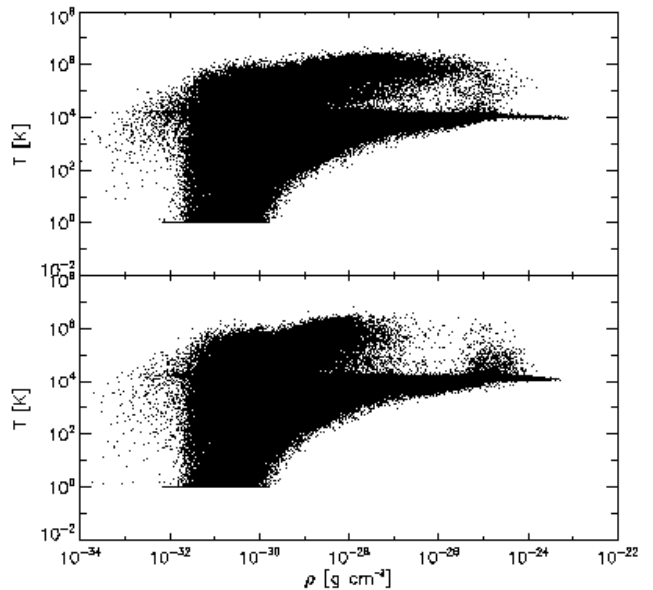}
\caption{Metal-dependent cooling clearly reduces the amount of
  hot gas in the halo (bottom panel) compared to the standard run (top
  panel). According to our definition for the size of the halo, gas with
  $\rho>9.29\times10^{-28}~\rm{g~cm^{-3}}$ belongs to the halo.}
\label{fig:metcool_impact}
\end{center}
\end{figure}

\begin{enumerate}
  \item A UV background \emph{(UV)}\\
    An external ultraviolet background is included in the Gadget2 cooling
    routine available to us from Volker Springel. It is a modified \citet{UV}
    spectrum with reionization at z$\sim$6 \citep{dave99}.
  \item Kinetic feedback \emph{(kin)}\\
    In order to stimulate the generation of winds by supernova explosions we test
    a model of splitting the supernova energy into a thermal and a kinetic
    feedback component. The ratio is governed by a parameter
    $\rm{f_v=E_{kin}/E_{SN}}$, which we
    set to 10\%. We follow a ``momentum'' approach as in \citet{NW93}, where
    the velocity of the neighboring gas particles is changed via
    \begin{equation}
      \Delta v_j = \sqrt{\frac{2 f_v E_{SN}(t)}{\sum{m_j}}}.
    \end{equation}
  \item Type Ia supernovae \emph{(SNI)}\\
    Type Ia supernovae arise from binary systems with a total mass of 3-16
    $\Msun$ and are particularly important for a delayed input of feedback
    energy as well as for metal enrichment since they are
    a main source for iron. We implement them following
    \citet{Scannapieco2006}, assuming a progenitor lifetime between 0.1 and 1
    Gyr. The exact time for energy (and metal) injection is chosen randomly
    from this range. The relative supernova rate of type Ia to type II is
    estimated from observations as 0.245 (\citet{cappellaro99}, for a Milky
    Way type galaxy (type Sbc) with $\rm{L_B(MW)}=2.3\times10^{10}\Lsun$). Each
    type Ia supernova is assumed to have an energy of $10^{51}$ ergs and we do
    not disable the cooling locally for these events
    \citep[following][]{stinson06}. 
  \item Mass return \emph{(met \& mSN)}\\
    In this model we return mass to neighboring gas particles in the same way
    as energy, smoothing it via the SPH Kernel. This should not have a
    big influence, but results in stars of different masses and therefore
    different amounts of returned feedback energy. This is also required to
    track metal enrichment which we do using oxygen and iron. For type II
    supernovae we use yields from \citet{WW95} taking into account the
    dependence on metallicity for the iron yields. To get more realistic
    abundances, one has to include type Ia supernovae, since they are mostly
    responsible for the iron production. For this we use yields from
    \citet{raiteri} which are based on \citet{thielemann86} and are
    independent of the progenitor mass. This is a separate model (\emph{mSN}),
    since the energy of type Ia supernovae is an additional factor.
  \item Metal-dependent cooling \emph{(metc)}\\
    The metal abundance in the gas has a strong influence on the strength of the
    cooling. At a temperature of $10^5$K, primordial gas will have a cooling
    time which is about 50 times longer than gas with [Fe/H]=0.5 \citep{SD93}. This results
    in a decreased amount of hot gas around the halo, as shown in
    Figure~\ref{fig:metcool_impact} in comparison to the standard run. Despite its
    importance, most current simulations do not include this effect
    (exceptions are \citet{Abadi2003}, \citet{Scannapieco2006}, \citet{okamoto05} and
    \citet{kawata05}). Based on our crude model of metal enrichment described
    above, we include metal-dependent cooling by interpolating the tables
    of \citet{SD93} which cover, besides the primordial case, the
    abundances [Fe/H]=$\{$-3.,-2.,-1.5,-1.,-0.5,0,0.5$\}$ and a temperature
    range of $4\leq \rm{log(T)}\leq 8.5$. For values outside these ranges, the
    respective extremes are used.
  \item Blastwave feedback \emph{(bl)}\\
    In our standard model using the SPH based method of smoothing energy
    and mass over neighboring gas particles with a variable-sized smoothing
    sphere, this size is computed so that the enclosed mass in the sphere is
    constant. This is numerically more sensible than our first
    attempt of using a fixed (albeit physically motivated) radius of 1.37 kpc
    h$^{-1}$. However, it is not very closely connected to the underlying physics it
    is supposed to represent. To improve on this, we implemented a version of
    the blastwave feedback suggested by \citet{stinson06}. In this model, the
    size of the smoothing sphere and also the time span for which cooling should
    be turned off are not free parameters, but calculated through the explicit
    blastwave solution based on \citet{chevalier74} and
    \citet{McKeeOstriker}. The blastwave radius is given by 
    \begin{equation}
      R_E = 10^{1.74}E_{51}^{0.32}n_{0}^{-0.16}\tilde{P}_{04}^{-0.2} pc
    \end{equation}
    and the timescale (i.e. the time until which cooling is disabled) is
    \begin{equation}
      t_E = 10^{5.92}E_{51}^{0.31}n_{0}^{0.27}\tilde{P}_{04}^{-0.64} yr
    \end{equation}
    with $E_{SN}=E_{51}10^{51}$ ergs and
    $\tilde{P}_{04}=10^{-4}P_0k^{-1}$. $P_0$ and $n_0$ are the ambient
    pressure and hydrogen density. We here use as time $t_E$ the time until
    the end of the snowplow phase, another option would be the time until the
    hot, low density shell survives. \citet{stinson06} found no significant difference
    between the two. Following \citet{stinson06} we distribute feedback energy
    to gas particles within $R_E$, but mass and metals (if turned on) within
    the original smoothing radius.  
\end{enumerate}

\subsection{Results}
\begin{figure}
\begin{center}
\includegraphics[width=0.5\textwidth]{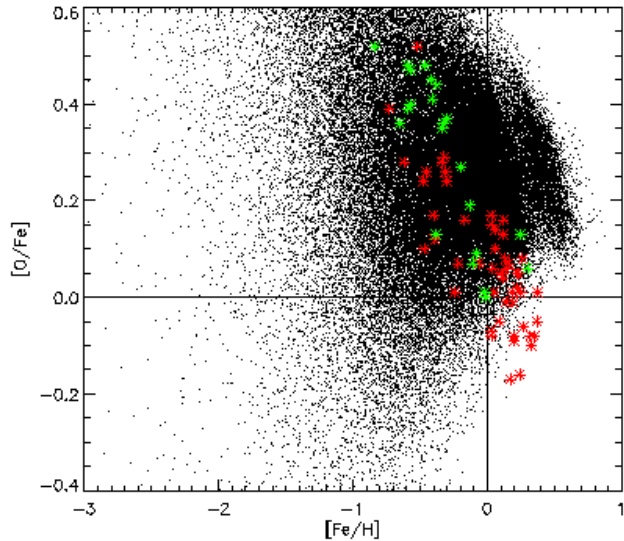}
\caption{Oxygen abundance in
  relation to iron for our stars in the standard model (black dots) 
  compared to observational data from the Milky Way by \citet{Bensby2004}. The
  green points are thick disk and the red points thin disk stars. Our simulation
  is in reasonable agreement with the former, though underabundant in iron.}
\label{fig:metals}
\end{center}
\end{figure}

\begin{figure}
\begin{center}
\includegraphics[width=0.5\textwidth]{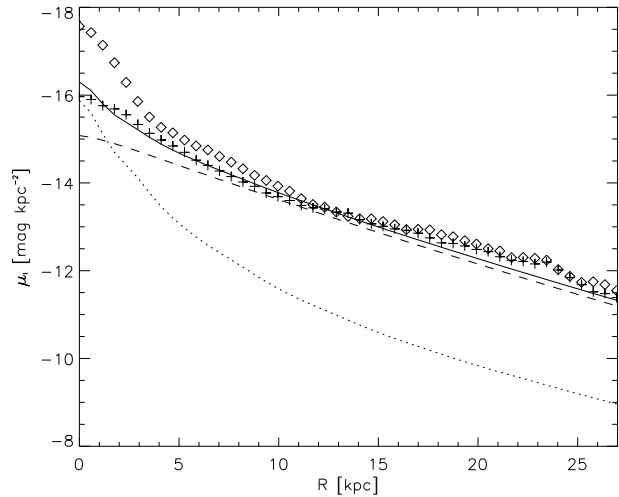}
\caption{The effect of including kinetic feedback on the surface brightness
  profile for the face-on projection of the galaxy. The diamonds show the profile of the standard run, the crosses the
  run with kinetic feedback. The solid line is the final fit combining a bulge
  (dotted line) and a disk (dashed line) component.}
\label{fig:surfaceB}
\end{center}
\end{figure}
From the visual impression of Figure~\ref{fig:physics_sky} it seems that all of
the additional physical effects except the blastwave method help to make the
galaxy younger and brighter. This is caused by a further
suppression of star formation early in the assembly history, leaving more gas
for later star formation. The blastwave method 
also causes such a suppression,
but in this case we additionally have a later
onset of star formation and a peak reduced to $9\Msun\,\rm{yr}^{-1}$ at
z$\sim$1.5. Still, despite much fewer stars and therefore less feedback
activity, this feedback is strong enough to result in the lowest galaxy mass
of all cases, with only 60\% of the mass in the standard case. Photometrically
the bulge is strongly reduced  and the galaxy also has the lowest luminosity
and current star formation rate of all models. Dynamically it is more a thick
disk system, though it efficiently retains its angular momentum. Overall, the
feedback in this case appears to be too strong, not only suppressing early, but also
current star formation too much and therefore preventing the formation of a
more dominant rotationally supported disk at z=0. This might be offset when
including other effects like metal-dependent cooling. Also, a better tuning of
the parameters in this model could improve the results, but that has to be left
to future work. \par
Looking now at the differences the other, more conventional individual effects
can cause, we first discuss the return of mass. 
As already expected, the differences here are small. The
galaxy is systematically slightly brighter due to a slightly higher star
formation rate which in turn is due to a higher cold gas fraction. The star
formation history is more characterized by bursts, but generally the characteristics
are comparable. \par
The model including type Ia supernovae 
is also similar. However, this is the only model
efficiently reducing the dynamical bulge. The bulge-to-disk ratio B/D is 0.18 compared to 0.43 in the
standard model. We attribute this mainly to a delayed peak in the star formation
history, at z$\sim$2.15 instead of z$\sim$2.6. The height of the peak is not really
reduced, but overall we have more gas and more younger stars which
consequently have formed already in the disk progenitor and therefore are
rotationally supported. They are not young enough to effectively brighten the
galaxy though, and photometrically the bulge is more prominent. As already
mentioned above, in the model combining type Ia
supernovae and mass return we also implemented some basic metal
tracking. The result is shown in Figure~\ref{fig:metals} for oxygen and iron
abundances in the stars. In comparison to the observations of Milky Way stars by
\citet{Bensby2004}, the agreement with thick disk stars (green points) is
reasonable, particularly in reproducing the trend (albeit with a large
scatter). Compared to the thin disk stars (red points) our results are a
little underabundant in iron. Since the model is rather crude this agreement
is satisfactory and also in overall agreement with results by \citet{scannapieco05}. Metal enrichment
is necessary for including metal-dependent cooling which will be discussed below.\par
The UV background 
shows influence on the star formation history only at lower redshift. The rather broad star formation peak
of the standard model is sharply decreased at z$\approx$2. In agreement
with results by \citet{Navarro1997} less gas cools into the disk which is
now surrounded by a halo of hot gas. The cold gas fraction is half of that in
the standard model. The disk is slightly smaller but also brighter and more
defined due to increased star formation particularly in the outer
parts. The disk has slightly less angular momentum than in the standard case,
and it is dynamically thicker at the expense of the thin rotationally
supported component. This again follows the trend discussed by
\citet{Navarro1997}, who concluded that a UV background makes it even
more difficult to form a high angular
momentum disk since it predominantly reduces the accretion of late infalling high angular momentum
gas. However, we find that including feedback can largely offset this negative
effect, resulting in a slightly, but not dramatically smaller angular momentum.\par
More interesting than the UV background is the model with kinetic feedback 
since it is the only one of our physical models (except the blastwave
model) with an almost pure 
exponential surface brightness profile for the face-on projection as shown in
Figure~\ref{fig:surfaceB}. The outer parts of the profile are very
similar to the standard run, but the bulge component in the inner 5 kpc is
strongly reduced. The star formation history is characterized by a deep drop at
$\rm{z}\sim2$, right after the first peak. The feedback generated from the
first stars is able to heat and blow out gas, permanently reducing the baryon
fraction within the virial radius to 0.11 (compared to a cosmic baryon
fraction of 0.167). Later, gas is not blown out of the halo anymore, but its
accretion onto the disk and therefore star formation in the disk is delayed, resulting
in a brighter disk and a reduced mass-to-light ratio from 2.18 to 1.93, using
the Chabrier IMF. The reduced star formation particularly at
early times leads to a reduction in the galaxy mass by 30\%, which is mostly a
bulge reduction. The reduced bulge
also leads to an increase in overall momentum making this our only run (besides the
blastwave model and the high resolution run) with no angular momentum
deficiency at all. The dynamical decomposition still shows a slightly reduced
rotationally supported component and an increase in the thick disk. \par 
Finally, including metal-dependent cooling 
basically offsets the feedback
effects completely. Star formation happens efficiently at all times,
especially very early resulting in a large peak of $\approx60\Msun\,\rm{yr}^{-1}$
at z$\approx$3. The feedback model is not able to regulate this at all. Star
formation is still happening efficiently at z=1 and a thin young disk is
present, but overall the system is strongly bulge dominated. Other effects are
much needed to control particularly the early star formation.

\section{Combining the physical models}
After studying each physical effect by itself, we now perform a series of runs
with what we call the "all in" model, a combination of the UV background,
kinetic feedback, type Ia supernovae and mass return. Acting in combination,
they of course will influence each other, which makes these runs more
realistic than the individual cases we discussed before. We first discuss the
result for the standard halo with standard resolution. Visually (see
Figure~\ref{fig:physics_sky}), the resulting galaxy is one of the youngest with a
strongly reduced bulge component. This is due to a very
efficient suppression of early star formation leaving more gas for later
accretion and leading to the highest current star formation rate of all our
models, 5.7$\Msun\,\rm{yr}^{-1}$. This suppression is a new effect due to the
combination of all the models, since, though all suppress star formation a
little at different times, none individually has such a large effect. This is shown in
\begin{figure}
\begin{center}
\includegraphics[width=0.5\textwidth]{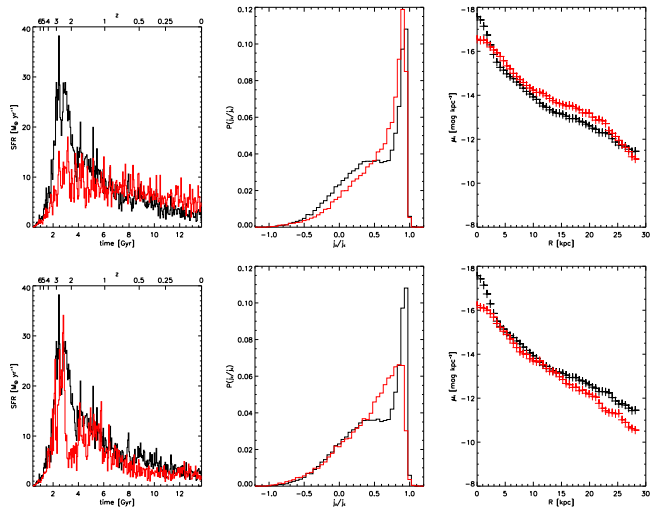}
\caption{A comparison of our
  standard model (black) with the "all in" model
  (red) in terms of star formation history (left column), dynamical rotational
  support (middle column) and I-band surface brightness profile (right
  column). The bottom row shows the same comparison, but now with the "all in"
  model including metal-dependent cooling ("allmc").} 
\label{fig:allin}
\end{center}
\end{figure}
Figure~\ref{fig:allin} in the left panel of the top row. The middle panel shows the
distribution of $j_z/j_c$ as a measure of rotational support. Clearly the
effect of the type Ia supernovae can be seen, shifting the emphasis from a
bulge to a thick disk and also slightly increasing the peak of the thin disk
with $j_z/j_c>0.85$. This thickening of the disk can also be seen from the
mock observational image of the edge on projection. Due to the quasi
elimination of the early star formation peak, 80\% of all stars in the final
galaxy are now
formed within the disk progenitor and therefore do not form a
spheroidal configuration. But particularly the stars older than about 7 Gyr do
not really remain in a tight thin disk, but their distribution thickens, leading to
the final result. \par
In the right panel of Figure~\ref{fig:allin}, we plot the surface brightness
profile for the face-on projection of the disk and again we can identify the
features contributed by the individual physical effects. The UV background is
responsible for the increased level outside of a
radius of about 18 kpc, with the sharp decline at the edge of the disk at
$\approx23$ kpc. The kinetic feedback reduces the bulge brightness in the inner region. This
is not as strong here as in Figure~\ref{fig:surfaceB}, which shows the model with kinetic
feedback as the only effect, since the combination of type Ia supernovae, mass
return and UV background together result in some overall brightening of the
whole galaxy. Overall, the galaxy characteristics are more realistic than
before. Each effect plays some role in this, though the kinetic feedback seems
to be the most important.

\subsection{Including metal-dependent cooling}
We now look at the "all in" run with metal dependent cooling included.
Results in comparison to the standard model
are shown in the bottom row of Figure~\ref{fig:allin}. As we have already seen
before in the run with standard feedback and metal-dependent cooling as an
individual effect, the
early star formation peak is strongly increased again.
Since metal enrichment proceeds quickly in the areas where star
formation can happen, the stronger cooling of gas with higher iron abundance
can take over quickly as well, making gas cool more efficiently, form stars
quicker and in turn speed up the enrichment. However, contrary to before, now
the combined physical effects can regulate
the increased star formation. The peak is narrow with a sharp drop (bottom
left plot of Figure~\ref{fig:allin}). The high early star formation also leads
to an increased fraction of stars formed in clumps again, now 37\% compared to
20\% without metal dependent cooling. In combination with
a lower star formation rate in the last 4 Gyr, this results in the strong reduction
of the thin disk peak in the dynamical decomposition shown in the lower middle
plot of Figure~\ref{fig:allin}. The thin disk is rather
unimportant dynamically with a mass fraction of only 17\% and it is more a thick disk-bulge
system. Photometrically, we have a fairly exponential profile with the kinetic
effect of reducing the bulge acting in combination with a reduction of
brightness in the outer regions of the disk. The UV background effects
evidently have been canceled out by the stronger cooling. From this we can
conclude that including other relevant physical effects, particularly kinetic
feedback, helps to overcome the additional problems of increased star
formation due to metal dependent cooling. However, a first high peak of star
formation cannot be prevented, and as a result it is much more difficult to
produce a dominant thin, rotationally supported disk at z=0. 

\subsection{The ``all in'' model in high resolution}
\begin{figure}
\begin{center}
\includegraphics[width=0.5\textwidth]{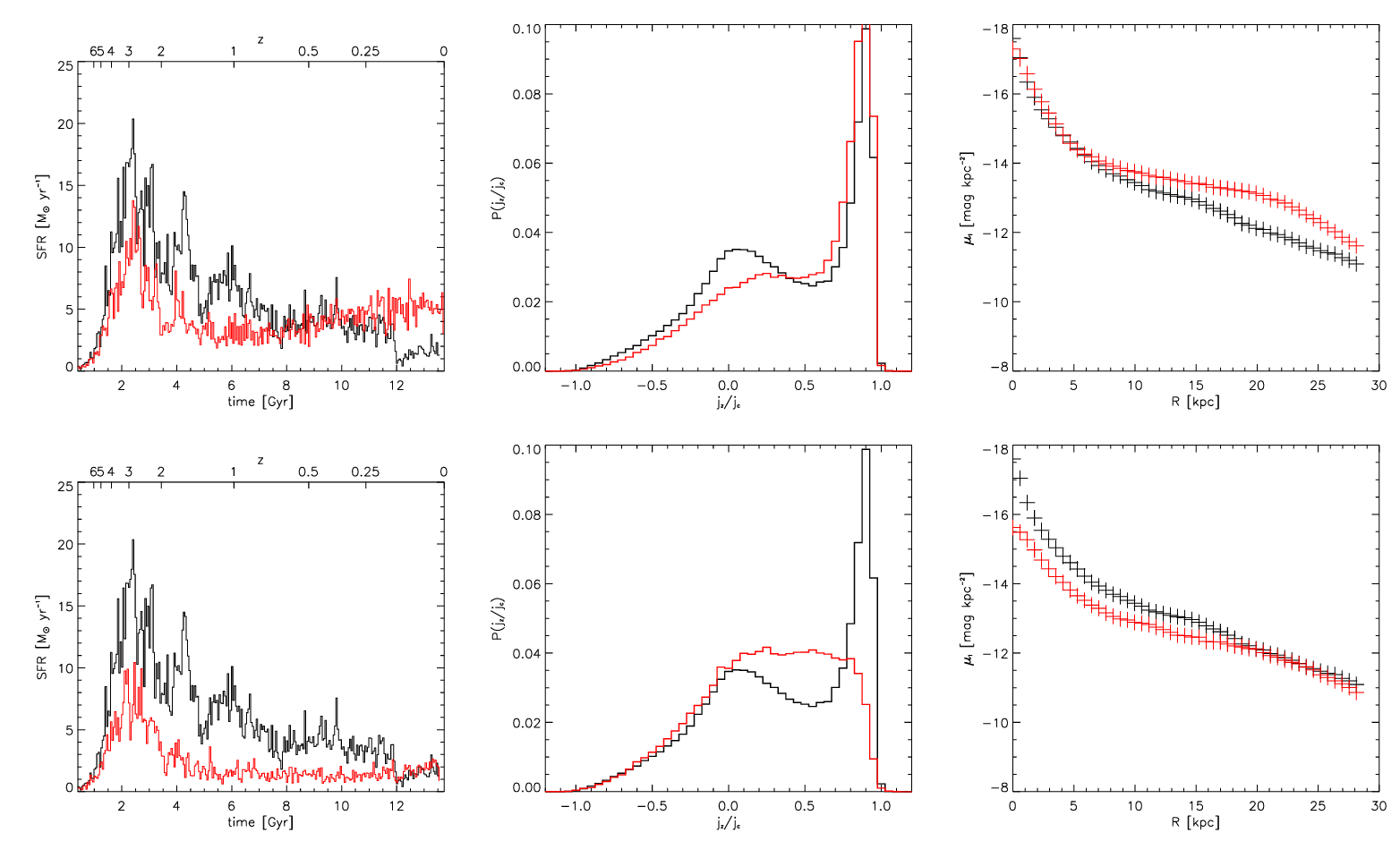}
\caption{Comparison of the standard (black lines) model with the ``all in''
  model (red lines) with 3\% (top panel) and 10\% (bottom panel) kinetic
  feedback for the standard halo in high resolution.}
\label{fig:2048allin}
\end{center}
\end{figure}
\begin{figure}
\begin{center}
\includegraphics[width=0.5\textwidth]{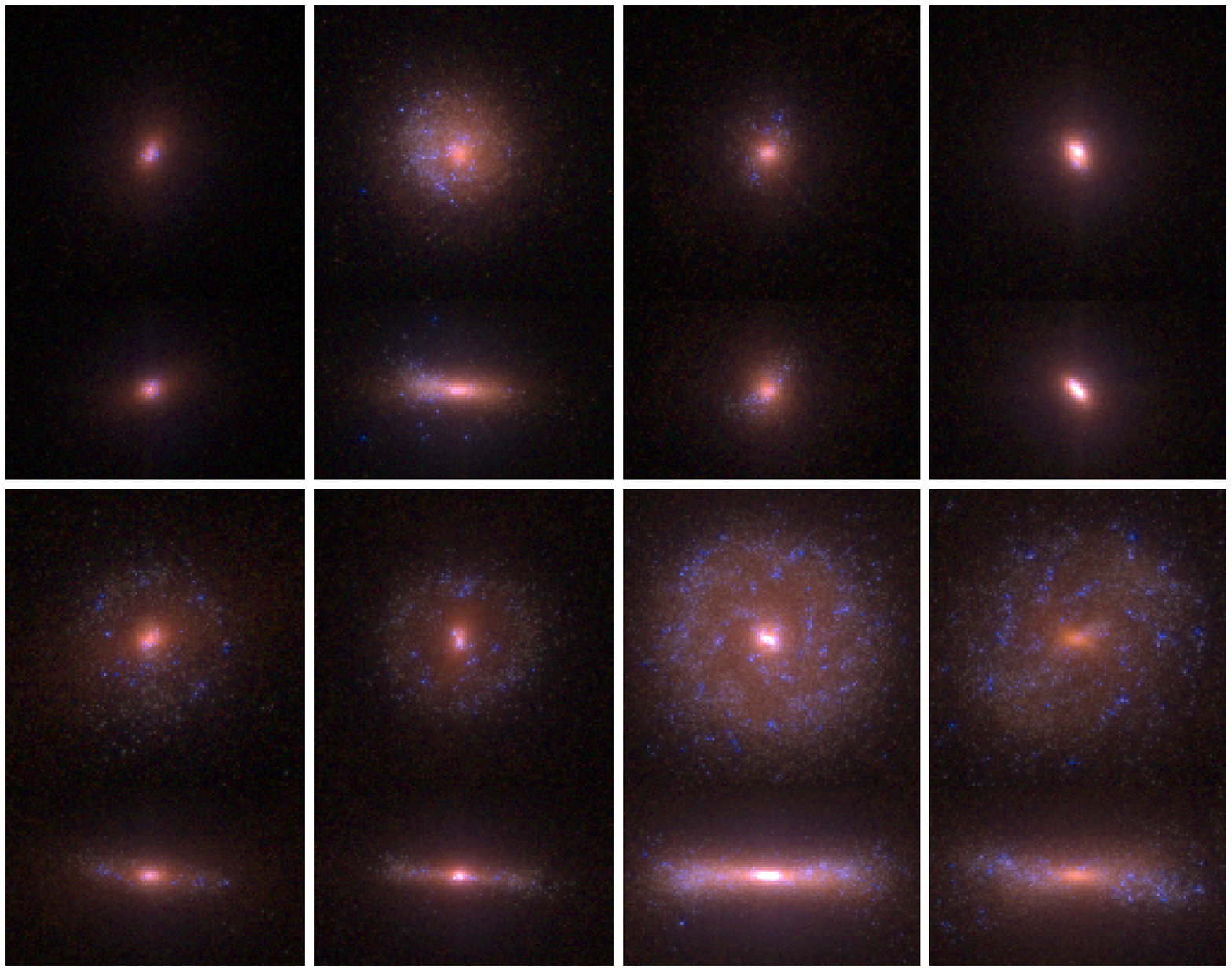}
\caption{Mock observations of the halos with different masses run with the
  ``all in'' model and 3\% kinetic feedback. The order is the same as in
  Figure~\ref{fig:diff_masses} except for the plot in the lower right panel
  which shows the standard halo with 10\% kinetic feedback in the ``all in''
  model.}
\label{fig:allin_pics}
\end{center}
\end{figure}
The "all in" model is not fully converged when applied to our higher resolution
halos. This can largely be attributed to the inclusion of kinetic feedback
which has a rather strong impact on the surrounding region of the stellar
particle. In the same way as in Figure~\ref{fig:allin}
Figure~\ref{fig:2048allin} shows the effect of two variations of the "all in" model for the
standard high resolution halo, MW$\_$hr. In the bottom row of plots, the model
is exactly the same as described in the previous section. The suppression of star
formation is extremely efficient, creating a system with a rather low mass of
$3.3\times10^{10}\,\Msun$, only 40\% of the stellar mass of the standard model
and 60\% of its luminosity. No thin disk is formed since there is only little
star formation after z$\approx$1.5. The galaxy has a rather low rotational
velocity and is quite thick. However, in agreement with the 
impact of kinetic feedback in the standard resolution run, the bulge component in the photometric
decomposition is reduced compared to the standard run. In order to reduce the
overly strong suppression of star formation and facilitate the formation of a
young thin disk, we reduced the amount of kinetic feedback from 10\% to
3\%. As can be seen in the upper row of Figure~\ref{fig:2048allin}, this is
quite successful for the dynamical decomposition. We find a bulge-to-disk
ratio of 0.57 while it is around 1 in the standard run and 1.26 in the
original "all in" model. However, 3\% is too little to drive out substantial
material from the central region of the galaxy and the flattening of the
surface brightness profile we saw in the standard resolution run with
10\% kinetic feedback disappears completely. The disk still is
substantially brightened in the outer parts leading to a lower mass-to-light
ratio (1.7 compared to 2.06) and a more prominent thin disk than in the standard run. The galaxy has no angular momentum
deficiency and, due to the higher luminosity at comparable rotational velocity,
the agreement with the Tully-Fisher relation is improved (though still not
perfect). Since this seems quite promising, we apply the low kinetic energy
"all in" model to all other halos with smaller masses than the standard
run. The resulting mock observations are shown in Figure~\ref{fig:allin_pics}. Generally, for the lower
mass halos the model is not very successful, with the exception of halo
DM$\_$hr2. All the others are rather spheroidal with low angular momentum and
higher mass-to-light ratios than before in the standard model. This shows that
the "all in" model even with lower kinetic energy is quite dependent on the
halo characteristics and also somewhat on resolution, and more work is needed
to investigate particularly the behavior of the kinetic feedback.

\section{The influence of the additional physics on the scaling relations}
\begin{figure}
\begin{center}
  \includegraphics[width=0.5\textwidth]{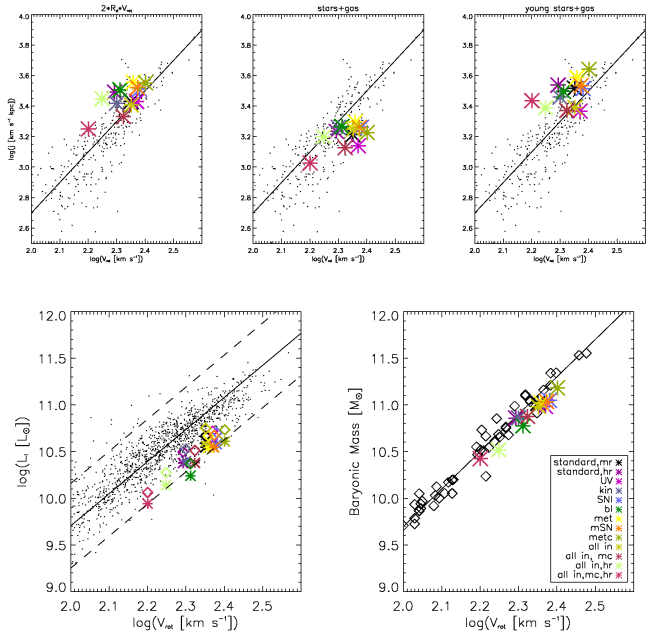}
\caption{Scaling relations for the different physical models as shown in Table~\ref{tab:physics}.}
\label{fig:scale_phys}
\end{center}
\end{figure}
\begin{figure}
\begin{center}
  \includegraphics[width=0.5\textwidth]{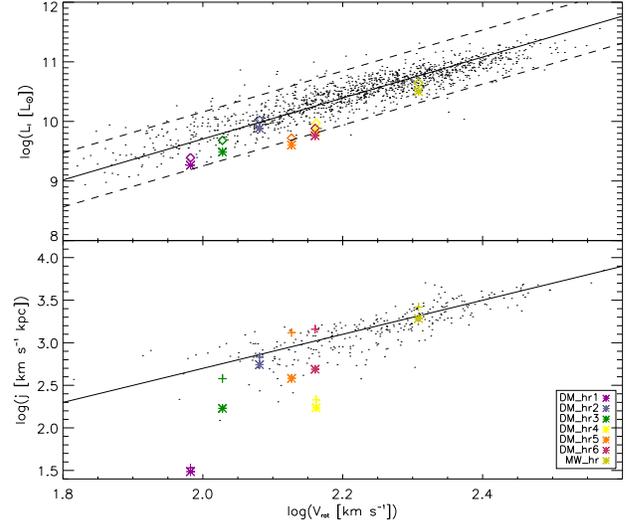}
\caption{Tully-Fisher and angular momentum relation for the different mass
  halos with the "all in" model with low kinetic feedback. In the upper panel,
  the star symbols are for a Salpeter IMF and the diamonds for a Chabrier IMF. The
  star symbols in the lower panel are the total angular momentum for the
  galaxy, the crosses for the disk of cold gas and  young stars.}
\label{fig:scale_allin}
\end{center}
\end{figure}
Figure~\ref{fig:scale_phys} shows the different physical models overplotted on
the observational scaling relations. Since these runs have all been done with
our standard halo, they scatter closely around the same value of rotational
velocity of around $200\,\rm{km}\,s^{-1}$. Most runs (with the exception of the model
with kinetic feedback, the blastwave model and the high resolution model) are slightly angular momentum deficient. This is mainly due to the old
stars in a more spheroidal configuration. In all cases there is a young
stellar disk which, in combination with the cold gas disk, fits well on the
angular momentum-size relation. This indicates again the difficulty of
comparing the actual angular momentum to the exponential disk estimator. We
conclude that the additional physical effects generally only mildly affect 
the angular momentum content, except for kinetic feedback, which
retains it much better than the others. However, as we have shown in our
angular momentum study in Paper 1, feedback itself is a necessary
ingredient to overcome the angular momentum problem.\par  For the Tully-Fisher (TF)
relation, we find a spread along the relation, but no single physical effect
can be tied to a strongly improved agreement with the observed relation. This
is consistent with the generally similar mass-to-light ratios, which in all
our models are between 2.5 and 3 for a Salpeter IMF and around 2 for a
Chabrier IMF. While the agreement is good with the baryonic TF relation
independent on the physical model used, there is some variation among the models with respect to the photometric TF relation. The lower the mass-to-light ratio,
the better the agreement with the TF relation. The standard resolution "all in"
model as well as the high resolution "all in" model with low kinetic feedback
both agree best. With M/L=1.7 the latter has the lowest overall mass-to-light
ratio. This leads us to the conclusion, that in general one physical effect
might induce small changes to the galaxy, but a big impact can only be
achieved with a combination. Our "all in" model, with its flaws, seems to be
promising in this respect.\par
The results of applying the "all in" model with low kinetic energy to the lower
mass halos is also in agreement with the conclusion drawn above. Their fit to
the scaling relations are shown in Figure~\ref{fig:scale_allin}. This plot
indicates a connection between angular momentum deficiency and poor agreement with
the photometric TF relation. The halos disagreeing with both relations also
tend to have high mass-to-light ratios. Only in two halos, the standard halo and DM$\_$hr2, does the "all
in" model result in an improved (in the latter case actually excellent)
agreement with all scaling relations. Both halos are the only ones close to
the observed Tully Fisher-relation (top panel of
Figure~\ref{fig:scale_allin}). They are also the only halos with no angular
momentum deficiency at all (already in the standard model and also in the "all
in" model; bottom panel of Figure~\ref{fig:scale_allin}). They have in
common the quietest merging histories of all our halos. They also have
rather stable star formation histories (halo DM$\_$hr2 basically lacks an
early star formation peak). If this is the prerequisite to form a reasonable disk galaxy, then again the question
is raised how the number of disk galaxies in the present day universe can be
reconciled with the high numbers of large mergers in a $\Lambda$CDM universe. 

\section{Discussion and Conclusions}
\begin{figure}
\begin{center}
  \includegraphics[width=0.5\textwidth]{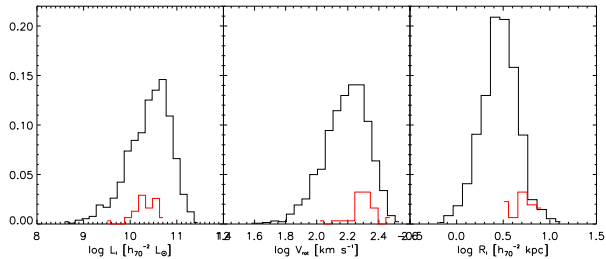}
\caption{The distribution of our galaxies (red curves) compared with observed
  data (black curves) from \citet{courteau07}, both normalized by their
  respective number of objects.}
\label{fig:typicalhalos}
\end{center}
\end{figure}
\begin{figure}
\begin{center}
\includegraphics[width=0.5\textwidth]{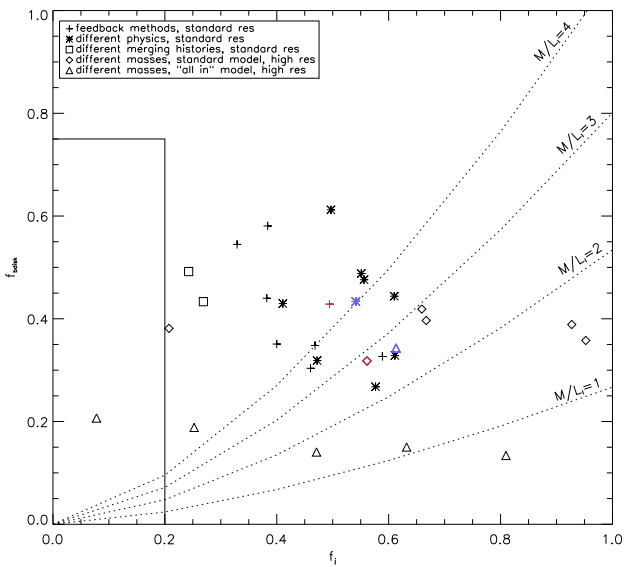}
\caption{Fraction of baryons assembled into the galaxy (f$_{\rm{bdisk}}$)
  vs. the ratio between the specific angular momentum of the disk and its
  surrounding halo (f$_{\rm{j}}$). The square marked by solid lines denotes
  the region populated by early simulated galaxies \citep{NavarroMS2000}. The red
  colored symbols show the standard halo with the standard model in standard
  (cross) and high (diamond) resolution. The blue colored symbols show the
  standard halo with the "all in" model for standard (star) and high
  (triangle) resolution.}
\label{fig:M_L}
\end{center}
\end{figure}
We have performed a detailed study of supernova feedback mechanisms in a Milky
Way-type halo, but also in a range of other halos spanning different masses
and different merging histories, using a medium and a high resolution. Out of a variety
of methods of distributing feedback energy we picked the one with the best
performance to be our standard model. The important feature for the
suppression of the formation of a large bulge was mainly to turn off cooling
locally to mimic a multiphase medium. The difference in the results between
the distribution methods themselves is comparatively small. The standard model is a model with an energy
output following an exponential law, local cooling turnoff and a variable
smoothing length for the supernova energy. It was then applied to halos with
different masses and merging histories and improved 
with additional relevant physical effects. The first question one might ask
is how typical are our resulting galaxies. In Figure~\ref{fig:typicalhalos},
we compare them to observed data \citep[from][]{courteau07} with respect to the most interesting
parameters of disk galaxies: the I band luminosity, the rotational velocity
and the I band disk scale length. Our galaxies are typical in terms of their
luminosity. The low outlier is our lowest mass halo. Their rotational
velocities are slightly high, but well within the distribution. The disk
scale lengths are all somewhat high. Even though we attempted to measure them
in the same way as an observer would, this is not straightforward 
and there may be systematic effects, especially due to the
rotation curves. While our rotation curves for the standard halo are mostly
flat enough that the exact measuring point does not matter, this is not always
the case for the lower mass halos, especially in the "all in" model.\par
From our study of different physical properties we can conclude that a
combination of type Ia supernovae, some fraction of kinetic feedback and a UV
background improves the result significantly towards a more realistic
disk. Particularly interesting is the kinetic feedback, which efficiently
blows out gas from the progenitor and reduces the importance of the central
bulge making the disk more exponential. However, this is a very harsh way of
changing the gas conditions and difficult to simply extend to higher resolution. In
our high resolution runs, we had to reduce the kinetic energy fraction from
10\% to 3\% to achieve reasonable results, despite the resulting loss of the
flattening effect in the surface brightness profile. Further investigation
seems necessary here to improve the stability of the model.\par
We also find that it is not enough to
simply reduce the early peak in star formation. This reduction leaves a lot of
gas for later accretion and star formation. It reduces the photometric
importance of the bulge, but also tends to thicken the disk. Dynamically, the
only way to decrease the bulge-to-disk ratio seems to be a delay of the star
formation peak, as in our case with the effect of type Ia supernovae, or to
prevent the gas to enter the disk at a later point in time. \par
Including metal dependent cooling has a significant impact and actually
raises the early star formation peak again, making it more difficult to form a
disk. In combination with the other effects, this can be controlled, but the
result is not quite satisfactory yet, indicating that a more complex model
might be necessary. However, our metal
enrichment is very basic. It seems to be in general agreement with other
simulations and also with observations, but more extensive testing would be
necessary to study enrichment processes. Still, we believe that including
metal dependent cooling is important to correctly evaluate the performance of
the feedback model.\par
We also briefly looked at a blastwave-type feedback. Our model is not as
sophisticated as in \citet{stinson06}, since for example we did not include
feedback energy output depending on the lifetime of the stars. In our
case this feedback is extremely strong and reduces overall star formation too
much to create a Milky Way-type galaxy. However, a fine-tuning of parameters could
lead to improvements and other effects might also balance this. It
would be particularly interesting to look at metal-dependent cooling in this
respect since stronger cooling at early times needs stronger feedback to
maintain a low star formation rate.\par
Our galaxies all fit well on the angular momentum-size relation, so they do
not suffer from the earlier problem of too compact, centrally concentrated
disks. This is primarily a result of the feedback, not an effect of a high
resolution (see Paper 1). The actual angular momentum content of gas and stars in our simulated galaxies
is mostly within the scatter of observed galaxies (with the exception of
some of our halos where no real disk was formed at all), but on the lower end
with about 60-80\% of the expected relation. The only runs with no angular
momentum deficiency are those with kinetic feedback and with the blastwave
model. This is alleviated when looking only at the angular momentum of the
actual disk consisting of young stars and cold gas. For the total angular
momentum of the galaxy we find a systematic offset when comparing directly to
the angular momentum calculated, as observers do, from $\rm{j=2R_dV_{rot}}$, which
disappears when looking only at the angular momentum of the disk made of young
stars and cold gas. A comparison between observations and simulations is
therefore only meaningful for clearly disk-dominated galaxies.
\subsection{The Tully-Fisher relation as a benchmark test}
\begin{figure}
\begin{center}
\includegraphics[width=0.5\textwidth]{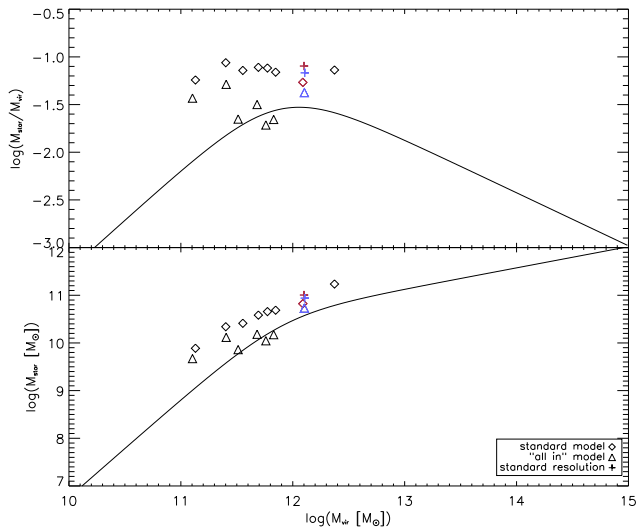}
\caption{The stellar-to-halo mass relation by \citet{Moster2009} in comparison with our
  simulated high resolution halos of different masses. The red and blue
  symbols indicate the standard, Milky Way mass halo, while the crosses are
  the same halo in standard resolution.}
\label{fig:SHM}
\end{center}
\end{figure}
\begin{figure*}
\centering
\subfigure[]
{
  \label{fig:fixed_ML}
  \includegraphics[width=0.45\textwidth]{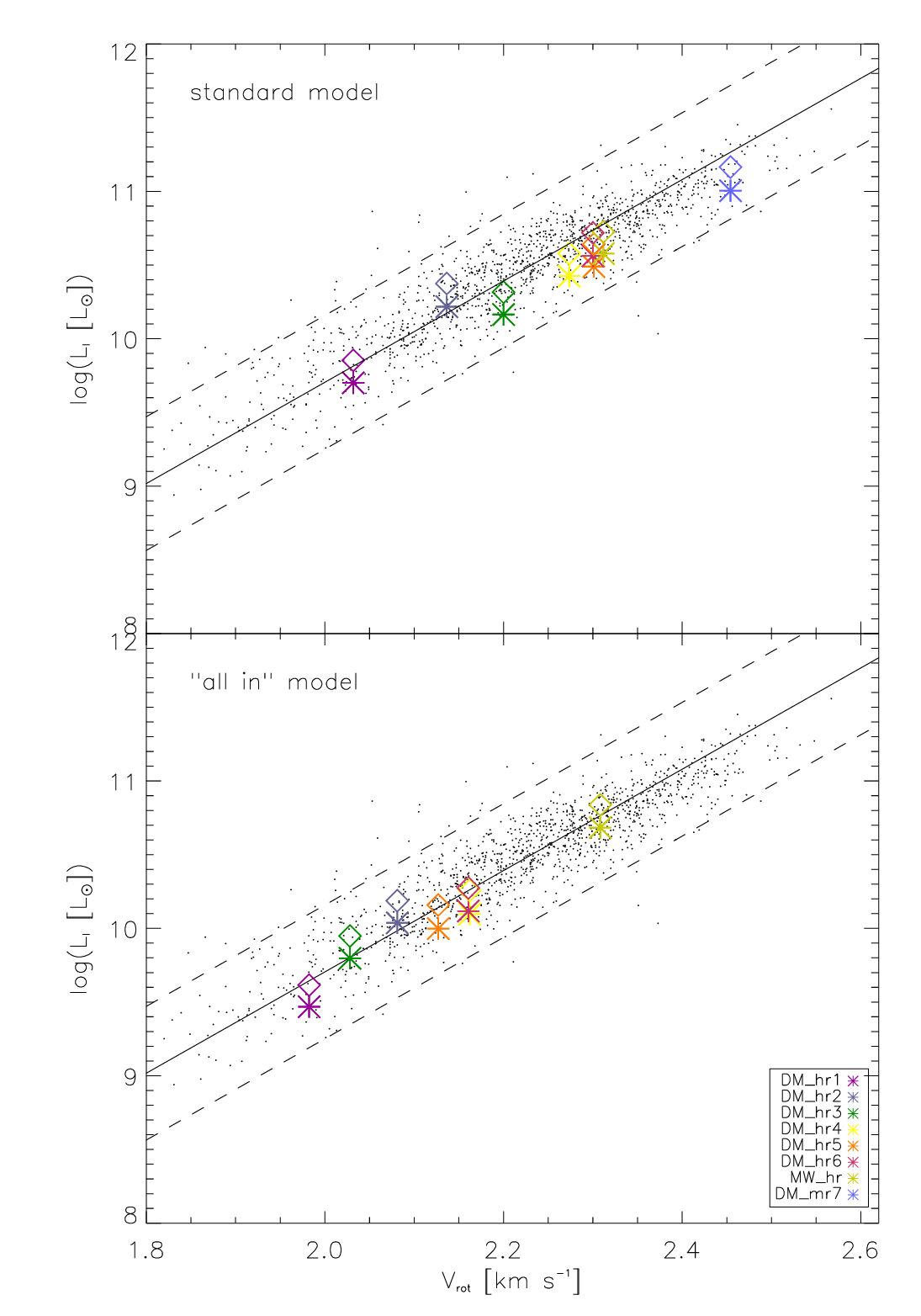}  
}
\subfigure[]
{
  \label{fig:fix_moster}
  \includegraphics[width=0.45\textwidth]{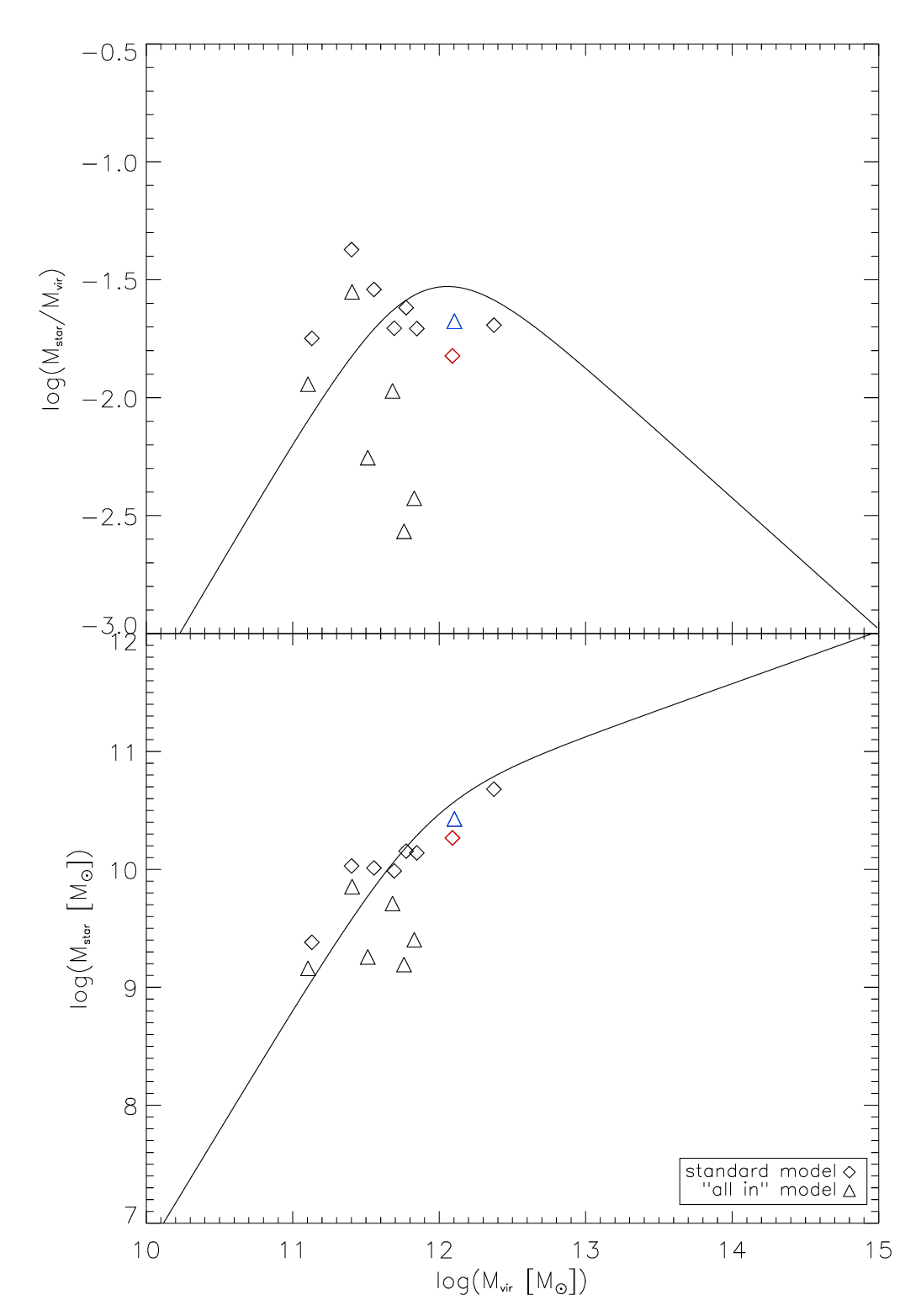}
}
\caption{Panel (a): Tully-Fisher relation with luminosities computed
  assuming that the bulges have the same mass-to-light ratio as the
  disk. Panel (b): Same as Figure~\ref{fig:SHM}, but with stellar masses
  excluding the bulge stars.}
\label{fig:fixplot}
\end{figure*}
The biggest challenge for a successful disk galaxy simulation remains to fit
the Tully-Fisher relation. We manage to fit the baryonic Tully-Fisher
relation quite well, as well as the slope for the photometric relation, but we
fail to reproduce the zero-point. From this we conclude that the structure of
our simulated galaxies is realistic. Including different physical effects 
scatters the galaxies along the photometric relation rather than improving the fit. The "all in"
model, with a combination of several effects, does result in an improvement over
the standard model. For a Salpeter IMF we improve from an offset factor of
$\approx$2.4 to 1.9, and for a Chabrier IMF from $\approx$1.8 to 1.4. However,
the agreement is still not satisfying. It is difficult to separate out the underlying
reason for the problem. It can either be that the luminosity is too low, or
that the rotational velocity is too high, or a combination of both. One
hint might come from the stellar mass-to-light ratios which in our case are
about 2 (based on a Chabrier IMF) with the "all in" runs of the standard halo
being improved to $\approx$1.6 both in standard and in high resolution. These 
increases again to about 2 when metal-dependent cooling is included. Further
information comes from Figure~\ref{fig:M_L} where we follow
\citet{NavarroMS2000} in plotting the fraction of baryons assembled into the
galaxy vs. the ratio between the specific angular momentum of the disk and its
surrounding dark matter halo. The simulated galaxies by \citet{NavarroMS2000}
were all placed within the squared region indicated by the solid lines. While
a large fraction of the baryons were assembled into galaxies, they had only
very little angular momentum. Our galaxies with their much higher angular
momentum content are much more distributed over the whole range of the plot,
but, with respect to the baryonic content, lie still preferably in the upper region,
above a fraction of 30\% of baryons being assembled into the disk. Exceptions are the lower mass halos with the "all in" model. Since
typical disk galaxies have a mass-to-light ratio of around 1-1.5, our
galaxies clearly contain too much mass or are too dim for the stellar mass
they have. From semi-analytical models we know that small baryon fractions are
needed to reach an agreement between the stellar mass function and the halo
mass function. In comparison with this our baryonic mass in the disk is too large.
This is confirmed by a comparison with the results by
\citet{Moster2009}, who derived a stellar-to-halo mass (SHM) relation
determined by the constraint to fit the observed SDSS stellar mass function
and correlation functions. This SHM relation is shown in
Figure~\ref{fig:SHM}, together with the results for our high resolution simulations of
different mass halos. The simulations with the standard model are shown with
diamonds and the ones with the "all in" model with triangles. The Milky
Way halo results are the red and blue symbols, respectively, while the crosses
show the corresponding result for the standard resolution halo. Our model
produces a systematic offset towards higher stellar masses which is worse for
the lowest mass halos. While the "all in" model is able to improve on that, we
know from above that these stars form a little bit too early and therefore are
too dim to fit the Tully-Fisher relation. As a side note it should be mentioned
that this curve does depend on the observed mass of the Milky Way dark matter
halo which is still under discussion. Recent results from SDSS and the
RAVE survey\footnote{The \textbf{Ra}dial \textbf{V}elocity \textbf{E}xperiment
is measures radial velocities, metallicities and abundance ratios for up to a
million stars in the Milky Way. See www.rave-survey.aip.de for more information.} point to a smaller mass \citep{Smith2007} which would decrease the gap
between the curve and our simulated data \citep{Xue2008}.\par
In Figure~\ref{fig:fixplot}, using ad-hoc assumptions, we explore ways on how to improve the agreement with the
Tully-Fisher relation and the SHM relation. Looking only at our disks made up of mostly
young stars, they have mass-to-light ratios of around 1.5 for a Salpeter and even 1
to 1.3 for a Chabrier IMF, which is in the range of observed disk
galaxies. For the different mass halos we therefore compute new total
luminosities assuming bulge luminosities based on the same mass-to-light ratio
as the disks. This is an attempt to model what the galaxies would look like if
the early star formation peak would be
delayed by some mechanism, possibly stronger stellar feedback. (We have a
small delay for the model including type Ia supernova feedback, which does
reduce the bulge, but the effect is not big enough.)  This
results in an excellent fit of the Tully-Fisher relation as
shown in Figure~\ref{fig:fixed_ML} for the standard model (top panel) and the
"all in" model (bottom panel). Since with this method only the luminosity is
adjusted, but the total stellar mass is not reduced, it would not result in an
improved agreement with the SHM relation. If a mechanism could be devised which not only would delay the
early star formation, but also prevent this gas from ever forming stars, and
therefore a bulge, for example by blowing it out efficiently or keeping it
hot, the mass would be reduced. Figure~\ref{fig:fix_moster} shows that in this
case, when only the stellar mass of the disk is taken into account, a good agreement
with the SHM relation is achieved for the standard model, though the shape
cannot be reproduced well. In the "all in" model, where the masses are lower,
the disk alone is not massive enough to fit the relation. An attempt to reproduce the Tully-Fisher relation under the
same assumption of taking out the bulge stars completely was not
successful. The effect of a smaller rotational velocity due to the reduced
mass is erased by the considerably reduced luminosity. This could be
because the disk mass-to-light ratios are smaller, but not small
enough. However, for the Chabrier IMF, they are close to 1 at least for
some of the galaxies. Another possible explanation is too much dark matter in
the halo dominating the rotation curve and maintaining a high rotational
velocity. This high dark matter concentration could also be a partial
explanation of the slightly low baryon fraction we find in some of our halos.\par
We conclude that even our more complex and realistic "all in" model is not
in complete agreement with the Tully-Fisher relation, though the galaxies are
within the observed scatter. An exception is halo DM$\_$hr2. Improvements could be
made in the implementation of already included effects (like a blastwave
approach) or with the addition of new effects, for example an efficient wind
model. However, from the discussion above, it seems that an improved fit to
both the Tully-Fisher and the SHM relation is possible only when the formation
of the bulge is largely prevented and the gas is kept outside the galaxy
permanently. None of the physical effects we tested was efficient enough to be
responsible for this. Our results indicate a possible further
complication in the high concentration of the dark matter halo, which could be
a basic problem of $\Lambda$CDM. We obtain the best results with two
halos with the most quiet merging histories, in line with the typical
assumptions for disk galaxies but in tension with $\Lambda$CDM's hierarchical
structure.\par In summary we find the question of the formation of realistic
disk galaxies in cosmological simulations a highly interconnected, complex
problem that is difficult to handle in a controlled manner. Particularly the
complex feedback models require a recalibration when increasing the
resolution. The answer to the pending problem of forming realistic disks does
not seem to be a single physical or numerical effect, but rather a combination
of many little effects whose interplay has to be scrutinized carefully and
systematically to find the solution.
\section*{Acknowledgements}
We thank Stefan Gottl\"ober and Gustavo Yepes for providing initial conditions
and Volker Springel for providing a version of \textsc{GADGET2} which included
radiative cooling. This work was funded through a grant by the German Research
Foundation (DFG) under STE 710/4 as part of the Priority Programme SPP1177
"Witnesses of Cosmic History: Formation and evolution of black holes, galaxies
and their environment".

\label{lastpage}

\end{document}